\documentstyle[prb,aps,twocolumn,epsfig,rotating]{revtex}
\begin{document}

\draft
\title{Dynamic scaling for 2D superconductors, Josephson 
junction arrays and superfluids}

\twocolumn[ 
\hsize\textwidth\columnwidth\hsize\csname@twocolumnfalse\endcsname 

\author{Stephen W.~Pierson,$^*$}
\address{Department of Physics, Worcester Polytechnic Institute 
(WPI), Worcester, MA 01609-2280} 
\author{Mark Friesen,$^{\dag}$}
\address{$^2$Physics Department, Purdue University, West Lafayette, IN 
47907-1396}    
\author{S.~M.~Ammirata,$^{\ddag}$}
\address{$^3$Department of Physics, Ohio State University, Columbus, 
Ohio 43210}  
\author{Jeffrey C. Hunnicutt and LeRoy A.~Gorham}
\address{Department of Physics, WPI, Worcester, MA 01609-2280} 

\date{\today}
\maketitle

\begin{abstract} 
The value of the dynamic critical exponent $z$ is studied for 
two-dimensional superconducting, superfluid, 
and Josephson Junction array systems in zero 
magnetic field via the Fisher-Fisher-Huse dynamic scaling. 
We find $z\simeq5.6\pm0.3$, a relatively large value indicative of 
non-diffusive dynamics. Universality of the scaling function is 
tested and confirmed for the thinnest samples.  We discuss 
the validity of the dynamic 
scaling analysis as well as the previous studies of the 
Kosterlitz-Thouless-Berezinskii transition in these
systems, the results of which seem to be consistent with 
simple diffusion ($z=2$). Further studies are discussed 
and encouraged.

\end{abstract} 
]
\narrowtext

\section{Introduction}
\label{sec:intro}

The dynamics of two-dimensional (2D) zero-field systems have been studied 
continually over the last few 
decades,\cite{ahns78,ahns80,taka80,mertens87,jonsson97,kim99} usually 
in the context of the
Kosterlitz-Thouless-Be\-re\-zin\-skii\cite{kt73,k74,b71} 
(KTB) transition. The dynamic critical exponent $z$ characterizes the
critical 
behavior of the dynamics of a transition. Despite the ongoing 
studies of KTB dynamics, the value of $z$ is usually not questioned 
to be anything but the value that describes simple diffusion: 
$z=2$. This paper, through an analysis of various transport data sets
from systems including superconductors (SC's), superfluids (SF's), 
and Josephson Junction Arrays (JJA's) using the dynamic scaling of
Fisher, Fisher and Huse,\cite{ffh} presents ample evidence that 
the value of $z$ in these systems is much higher: $z\simeq 5.6$. 
The purpose of this 
paper is to convince the reader that despite the many previous 
reports consistent with $z=2$, the question of the value of $z$ is 
still an open one. Perhaps the single element that distinguishes 
this analysis from past analyses is that $z$ is not taken 
{\it a priori} to have a value $2$. In the dynamic scaling analysis, 
varying $z$ to optimize the description of the data is standard.

This paper is organized as follows. We begin in this section 
with a general description of the KTB transition, previous 
scaling attempts of $I$-$V$ data, and previous findings of anomalous vortex 
diffusion. In Section \ref{sec:background}, we will discuss 
the various length scales whose competition results 
in the interesting critical behavior of this transition. In 
Section \ref{sec:results}, we present scaling results on SC's, 
JJA's, and SF's and also check for universality of the scaled 
data. We discuss the validity of the dynamic scaling approach 
as well as that of the conventional approach in Section 
\ref{sec:discussion}. We summarize the paper in Section \ref{sec:summary}.

The KTB transition is driven by the unbinding of vortex pairs. Below the 
transition temperature $T_{KTB}$ vortices are thermally induced and can 
only be excited in pairs that have a finite energy and not as ``free" 
vortices, which have an infinite energy in an infinite system. As the 
temperature is increased, the number and size of the vortex pairs 
increase and these pairs screen one another's interactions. At the 
transition temperature, the vortex pairs start to unbind and  
free vortices are formed. It is the largest pairs that unbind first, 
leaving a finite density of smaller pairs above the transition 
temperature. The size of the largest pairs decrease as one goes 
further above $T_{KTB}$. As we review below, for $T>T_{KTB}$, the free
vortices result in an ohmic $I$-$V$ curve at low currents while the smaller 
pairs result in a non-linear $I$-$V$ relationship at larger currents.

Scaling techniques applied to zero-field $I$-$V$ data 
from SC's or JJA's have been reported in the past but without
finding the results reported here. In Ref.~\onlinecite{wolf79}, Wolf {\it et
al.} scaled their $I$-$V$ data from granular 30$\AA$ thick films. 
There, however, the dynamic universality class was not explicitly 
studied and no value of $z$ was determined. D.~C.~Harris 
{\it et al.}\cite{harris91} reported $z=2$ in a dynamic scaling analysis 
of Josephson junction array $I$-$V$ data, although it appears that they 
did not allow $z$ to vary. We address that data in Section
\ref{sec:jjascaling}. 
L.~Miu {\it et al.}\cite{miu95} mention peripherally a dynamic scaling
analysis on BSSCO (2223) in which it is found that $z\simeq 4.5$ (assuming a 2D
system). However, those results were not pursued further by those authors. In the case
of superfluids, Brada {\it et al.}\cite{brada93} have performed a finite 
size scaling analysis (in contrast to a dynamic scaling analysis) 
of their frequency dependent superfluid density 
and dissipation data from helium films. They however did not include a 
study of the value of $z$ in their work.

There has been experimental work that points toward anomalous
vortex diffusion. Th\'eron {\it et al.}\cite{theron93} used impedance 
measurements on weakly 
frustrated JJA's to provide evidence of non-conventional vortex dynamics. 
Unfortunately, this work did not probe the value of $z$. There has also been
theoretical\cite{bormann} and simulational\cite{kim99,noteminn} work that 
indicates anomalous vortex diffusion have been made, but none conclude
that $z>2$ at $T=T_{KTB}$. In this work, in order to limit the scope 
of this already long paper, we will not discuss the theoretical and 
simulational work, but concentrate only on experimental data. 

In comparing the ``conventional'' approach with the dynamic scaling approach, 
we will make use of many of the formulas derived from the former. However, 
because the dynamic scaling approach indicates that the dynamics are 
non-diffusive, the assumptions used to derive the formulas for the 
dynamics in the conventional approach may be incorrect. For that 
reason, we will use those formulas only to address the validity 
of the conventional approach and not the validity of the dynamic 
scaling. 

It is useful for us to clarify our use of the phrase ``conventional
approach." Because the dynamic scaling has only been used sparingly in analyses of the
type of data that we look at here, we view this approach as non-conventional. We 
will  therefore refer to any other approach to analyzing this
data as ``conventional" or traditional.

The value of $z$ that is found here runs contrary to some of the  
conclusions of previous studies. However, we stress that our results are not 
inconsistent with all of the results in the literature. In particular, the 
dynamic scaling results presented here do not contradict (or even pertain to)
the conventional findings for static behavior of the KTB transition. 
Further, our results differ from only some of the experimental studies of
the dynamics. This will be discussed in Section \ref{sec:validityCA} but we
mention some of those now. If the measurement does not involve $z$ or if the 
measurement does involve $z$ but 
can measure only the product $bz$ (where $b$ is a material-dependent constant
that enters through the correlation length [see Eq.~(\ref{xi+})]), then our
findings do not contradict those measurements. An example of the first type of
measurement are  ``static'' kinetic inductance measurements.\cite{fiory83} Examples of
the second class of measurement are the resistance, noise spectrum 
measurements,\cite{shaw96} and helium torsion measurements.\cite{bishop78,bishop80}

\section{The KTB transition in 2D superconducting systems: background}
\label{sec:background}
In this paper, we propose an interpretation of transport data on 
superconductors, superfluids and Josephson Junction arrays that 
is very different than that which has been accepted for the 
last twenty years. In order to judge the two approaches, a 
thorough understanding of the KTB transition and critical 
behavior as well as the approaches used to study them is needed. 
In this section, we will 
review this background, directing our discussion primarily at 
superconducting systems. (For discussion on JJA's or superfluids, 
see the reviews listed below.) We will discuss the 
various relevant length scales, the criteria for a phase transition, 
and the approaches one can take to study the KTB dynamic critical behavior.
We don't intend our review to be comprehensive and refer the reader to 
any of the many excellent reviews.\cite{halperin79,mooij84,minnhagen87}

\subsection{Length Scales}
\label{sec:lengths}

The competition of the length scales\cite{hebard} in the system 
determines the critical
behavior of the KTB transition. For this reason it is important to
review each of them. One can subdivide the length scales of the 
system into two categories: intrinsic and extrinsic. 
By extrinsic, we mean those length scales that are determined by an 
applied current or magnetic 
field.

{\centerline{\it Intrinsic}}

The intrinsic length scales include the vortex correlation length 
$\xi(T)$, the 2D penetration 
depth $\lambda_{2D}=2\lambda^2/d$ (where $\lambda$ is the London 
penetration depth and $d$ is 
the sample thickness), and sample size.

There are three important aspects of the correlation length that we 
discuss here. The first is its distinctive temperature dependence 
for temperatures above the transition temperature $T_{KTB}$:\cite{k74}
\begin{equation}
\label{xi+}
\xi_+(T) \propto \exp[\sqrt{b/(T/T_{KTB}-1)}]
\end{equation}
where $b$ is a non-universal constant. This unique temperature 
dependence is in contrast to the common power-law dependence one 
finds, for example, in Ginzburg-Landau theory.

The second aspect is the behavior of $\xi_-(T)$ below $T_{KTB}$. 
Because the susceptibility 
below the transition temperature is infinite, Kosterlitz originally 
defined $\xi_-(T)$ to be 
infinite. Based on the critical behavior of the dielectric constant, 
Ambegaokar {\it et al.} 
(AHNS)\cite{ahns78} defined a finite diverging correlation length for 
$T<T_{KTB}$. The two 
results do not contradict one another since they have different 
meaning. The AHNS correlation 
length for $T<T_{KTB}$ can be thought of as the size of the largest 
vortex pairs.\cite{simkin97} 
Ambegaokar {\it et al.}\cite{ahns80} estimate that the $\xi_-(T)$ 
has a smaller magnitude than 
$\xi_+(T)$: 
\begin{equation}
\label{xi-}
\xi_-(T) \propto \exp[\sqrt{b/2\pi(1-T/T_{KTB})}]
\end{equation}
In this paper, we will take $\xi_-(T)$ to represent the 
size of the largest pairs.

The third aspect of the correlation length that is important 
in this paper is its behavior 
in an applied current $I$. The effect of an applied current 
is to unbind vortex pairs down 
to zero temperature and therefore to destroy the phase 
transition. As a result, the 
correlation length no longer diverges for finite $I$ at $T_{KTB}$ and 
has the following behavior:
\begin{equation}
\label{xiI}
\xi_\pm(T,I) \propto \frac {T}{I} f[I\xi_\pm(T,I=0)/T],
\end{equation}
where $f$ is a non-singular function.

The next two intrinsic length scales are associated with 
finite size effects and give a 
single vortex a finite ``bare" energy. The first, which does 
not apply to superfluid Helium, 
is the 2D penetration depth $\lambda_{2D}=2\lambda^2/d$. At 
distances less than this length 
from the vortex core, the superfluid velocity goes as $1/r$. 
Beyond this length, the superfluid velocity decreases as 
$1/r^2$.\cite{pearl64} In a ``perfect" ($d=0$) superconductor, 
$\lambda_{2D}=\infty$ and the superconductor behaves as a 
superfluid would.  The second finite size length is sample 
size, which is the smaller of the sample width $W$ and length 
$L$. (Typically $W\ll L$.) The energy of a 
free vortex is,
\begin{equation}
\label{freevortex}
E_{FV}=[q^2/2]\ln (L_{fs}/\xi_0) +E_c
\end{equation}
where $\xi_0$ is the size of the vortex core and the finite size 
\begin{equation}
\label{Lfs}
L_{fs}=\min[\lambda_{2D},W]. 
\end{equation}
The vortex interaction strength $q^2$ and the core energy $E_c$ depend 
on the system (SC, SF, of JJA).  A finite energy for a single vortex 
means that there will
be free vortices below the transition temperature, which in turn 
precludes a true phase transition as we will discuss in the next section.  

\centerline{{\it Extrinsic}}

Three extrinsic length scales characterize the application of 
$dc$ and $ac$ electric and magnetic fields. The first is the length scale 
$r_c$,\cite{ahns80,goldman83} which is the length scale probed
by a dc applied current. For a superconducting film, the energy of a 
vortex pair with separation $R$ is\cite{minnhagen87,hn79}
\begin{equation}
\label{pairenergy}
E(R)=[\pi n^{2D}_s\hbar^2/2m] \ln (R/\xi_0)-\pi\hbar I R d/e A +2E_c
\end{equation}
where the vortex interaction strength has been expressed in terms of the
superconducting parameters $q^2=\pi n^{2D}_s\hbar^2/2m$ ($n^{2D}_s=n_sd$ 
is the areal superfluid density, $n_s$ 
is the superfluid density, and $m$ is the mass of a free electron,) 
and $A(=Wd)$ is the cross-sectional area through which the 
current $I$ flows. For small separations ($R<r_c$), the 2D, 
logarithmic term dominates and the interaction is attractive. 
For $R>r_c$, the linear, current-induced term dominates and so 
the interaction is repulsive. As a result, the interaction energy 
peaks at
\begin{equation}
\label{rc}
r_c=4k_BT_{KTB} W e/\pi\hbar I
\end{equation}
where we have written the interaction strength in terms of the transition 
temperature,\cite{hn79} 
$\pi n^{2D}_s\hbar^2/2m=4k_BT_{KTB}$. The non-linear $I$-$V$ relationship 
originates in thermally-activated hopping over this barrier at a rate 
$\Gamma$ which depends
on the value of the vortex pair energy $E(R)$ at this separation:
$\Gamma\propto 
\exp[E(r_c)/k_BT]$. Therefore, the dc $I$-$V$ curves probe length scales of 
O($r_c$).\cite{ahns80,goldman83}

In $ac$ measurements (e.g. kinetic inductance) with circular frequency
$\omega$, the 
probing length is the diffusion length, $r_\omega=(14D/\omega)^{1/2}$, 
where D is the vortex diffusion constant.\cite{ahns78,ahns80,at79} 
This result is derived by analyzing the linear response of the dielectric 
constant. It should be pointed out that the two quantities, $r_c$ and
$r_\omega$, 
do not compete per se with the other length scale but rather
indicate the length scale being probed.

The final extrinsic length scale that we mention is due to an applied 
magnetic field and characterized by the average distance between 
field-induced vortices 
$l_B\simeq(\Phi_0/B)^{1/2}$ where $\Phi_0$ is the superconducting flux 
quantum and $B$ is the magnetic induction. The field-induced vortices 
are ``free" and present at all temperatures, which precludes a true 
phase transition.

\subsection{Existence of the phase transition?}
\label{sec:transition}

Because of $\lambda_{2D}$, Kosterlitz and Thouless 
originally wrote that this critical behavior would not apply to 
superconductors.\cite{kt73} It was later realized\cite{hn79,beasley79} 
that in practice, $\lambda_{2D}$ can be larger than the system 
size and so superconducting films should not behave much differently 
than superfluid films. Whether $\lambda_{2D}$ is larger than the 
system size or vice versa, there will be a finite density of free vortices
below the 
transition temperature. The density of free vortices will 
be\cite{hn79,herbert98}
\begin{equation}
\label{density}
n_F\propto L_{fs}^{q^2/2k_BT}
\end{equation} 
where $q^2$ here is the renormalized vortex interaction strength and, again, 
$L_{fs}=\min [\lambda_{2D},W]$. Note that system size and 
$\lambda_{2D}$ enter Eq.~(\ref{density}) in the same manner, implying 
that one cannot tune the ratio of $\lambda_{2D}$ to $W$ to observe 
the transition. This is contrary to a 
perception in the literature\cite{herbert98,lobb83,lobb96} that the 
transition can be observed if $\lambda_{2D}>W$. This begs the question 
of why one should see critical behavior at all if there is no true 
phase transition in finite size systems. The answer of course is that 
critical behavior can be seen if there is a diverging length. 
This occurs in the KTB system, provided that $\xi_\pm(T)<L_{fs}$.

\subsection{Approaches to studying KTB dynamics in superconductors}
\label{sec:approaches}
A variety of approaches can be used to study KTB dynamic behavior. 
Here we review the two main approaches used to study the dynamics of
superconducting and JJA films and that therefore determine a 
value for the dynamic critical exponent $z$. Brief mention of 
the methods used for superfluid helium systems will be made in 
Sections \ref{sec:sfscaling} and \ref{sec:inductance}. There are other 
approaches for investigating KTB behavior that don't uniquely determine 
a value for $z$; we will not review here but we will discuss 
them  in Section \ref{sec:validityCA} in the context of 
previous evidence that $z=2$. 

In this section, we will first review the conventional results and 
its derivation. We will then generalize these formula to a general
value of $z$. The dynamic scaling analysis will then be introduced
and finally, the connections between the two approaches will be
discussed. 

\subsubsection{Conventional approach}
\label{sec:approaches1}

In the first, more ``conventional'' approach, the 
current-voltage ($I$-$V$) isotherms are measured 
and analyzed in terms of their $I\rightarrow 0$ 
limit:\cite{minnhagen87,hn79} 
\begin{equation}
\label{iv}
V/I\propto I^{\alpha(T)-1}.
\end{equation}
The signature of a KTB transition is a jump from non-linear behavior
below the transition temperature to ohmic above the $T_{KTB}$:
\begin{equation}
\label{RT2}
V/I\equiv R(T)  \propto \exp[-2\sqrt{b/(T/T_{KTB}-1)}]
\end{equation}
where $b$ is a non-universal constant. [Minnhagen\cite{minnhagen87} 
has generalized this to take into account the underlying 
superfluid. See Eq.~(\ref{RTmin}).] In particular, the exponent 
$\alpha(T)$ will decrease linearly with increasing temperature 
until the transition temperature is reached at which point 
$\alpha(T)$ will, in the $I\rightarrow 0$ limit, jump from 
3 to 1\cite{hn79} because of the ``universal jump'' in the 
superfluid density.\cite{nelson77} Indeed, the condition 
$\alpha(T=T_{KTB})=3$ is commonly used to determine the 
transition temperature. 

Eq.~(\ref{iv}) is derived by determining $R(T,I)$ above and 
below $T_{KTB}$ and using $V=IR(T,I)$. This derivation is well 
documented\cite{goldman83,hn79,pierson97} 
but we will highlight the key points. To find $R(T,I)$, the 
density of free vortices $n_f$ must be determined. 

Below the transition temperature, free vortices in the limit 
of a weak current are produced by thermal activation over the 
barrier in $E(r_c)$ as mentioned above [below Eq.~(\ref{pairenergy})]. 
This is done using the kinetic equation for the rate of change in the
number of free vortices:
\begin{equation}
\label{kinetic}
dn_f/dt=\Gamma(T,I)-n_f^2.
\end{equation}
As mentioned above, $\Gamma(T,I)$ is the rate at which vortex 
pairs are unbound [$\Gamma(T,I)\propto \exp\{-E(r_c)/k_BT\}$.] 
The second term in Eq.~(\ref{kinetic}) takes into account free 
vortices combining to form pairs. In steady state, 
\begin{equation}
\label{nfGamma}
n_F=\Gamma(T,I)^{1/2}\propto I^{q^2/2k_BT},
\end{equation}

Above the transition temperature, 
\begin{equation}
\label{nfxi}
n_f\propto \xi_+^{-2},
\end{equation}
since $\xi_+$ is the average distance between free vortices.

The final step in determining Eqs.~(\ref{iv}) and (\ref{RT2}) is 
to relate $R(T,I)$ to 
$n_f$:\cite{goldman83,hn79,huberman78} 
\begin{equation}
\label{Rnf2}
R\propto n_f.
\end{equation}
Substituting Eq.~(\ref{nfGamma}) and Eq.~(\ref{nfxi}) into Eq.~(\ref{Rnf2}) 
and using $V=IR(T,I)$ one arrives at Eqs.~(\ref{iv}) and (\ref{RT2})
respectively. 
Kadin {\it et al.}\cite{goldman83} have made extensions 
of this work to finite current. Many workers in the field however take 
Eq.~(\ref{iv}) to be valid over wide ranges of $I$. 

Eq.~(\ref{Rnf2}) is based on the Bardeen-Stephen flux-flow 
formula,\cite{bs65} (as stated by subsequent 
authors.\cite{fiory83,minnhagen87}) For subsequent discussion,
it will be useful to outline the derivation of this equation,\cite{hn79} 
whose starting point is the electrodynamic Josephson relation, 
\begin{equation}
V=[\hbar/2e]d\Delta \phi/dt,
\label{edJ}
\end{equation}
where $\Delta \phi$ is the change in the phase of the superconducting 
order parameter across the width of the sample.  
$d\Delta \phi/dt$ is proportional to the number 
of vortices that cross the width of the sample per unit time, 
\begin{equation}
\left\vert d\Delta \phi/dt\right\vert = 2\pi L n_F \left\vert v_D\right\vert,
\label{Delphi}
\end{equation}
where $v_D$ is the vortex drift velocity. 
Finally, one assumes
\begin{equation}
v_D= \mu\pi\hbar I/eA,
\label{vD}
\end{equation}
where $\mu$ is the vortex mobility and $e$ is the electron charge. 
Because the vortex mobility is taken
to be local and therefore independent of $n_F$, Eqs.~(\ref{edJ})-(\ref{vD}) 
result in Eq.~(\ref{Rnf2}): $R\equiv V/I\propto n_F$. 

The linear relationship between $R(T)$ and $n_f$ in Eq.~(\ref{Rnf2}) presumes 
single vortex diffusion. If one were to allow more complicated 
dynamic or critical behavior, then Eq.~(\ref{Rnf2}) must be modified.  
The general expression for $R(T)$ in the critical region depends upon 
the dynamic exponent $z$:
\begin{equation}
\label{Rnf}
R\propto \xi^{-z}\propto n_f^{z/2}.
\end{equation}
Of course the two expressions are identical for $z=2$. 
Using Eq.~(\ref{Rnf}) in place of $R\propto n_f$ in the conventional
derivation,\cite{fiory83,goldman83,hn79} one
would find that $\alpha(T)$ will jump from $z+1$ to 1 in the $I\rightarrow 0$
limit and that \begin{equation}
\label{RT} R(T)  \propto \exp[-z\sqrt{b/(T/T_{KTB}-1)}]. \end{equation} We
will
discuss the ramifications of using
$z\not=2$ in Eq.~(\ref{Rnf}) in Section \ref{sec:approaches3}.

\subsubsection{Dynamic scaling approach}
\label{sec:approaches2}

Dynamical scaling is motivated by the observation of critical slowing down
near a continuous transition.  This phenomenon is well established for
the KTB transition, and is marked by the divergence of the relaxation
time scale, $\tau$.  The dynamic scaling hypothesis asserts that critical
slowing down is related crucially to the divergence of the static
correlation length:\cite{goldenfeld} $\tau \propto \xi^z$.  

In general, many types of dynamics may be associated with a particular static 
universality class, and these should fall into distinct dynamical
universality classes.\cite{hohenberg} For two-dimensional
systems, including SC's, SF's, and JJA's, the conventional KTB dynamical
theory\cite{ahns78,ahns80} is consistent with model A universal
dynamics,\cite{hohenberg} and $z=2$.  For bulk superconductors however,
the correct dynamical universality class is presently unclear, and seems 
{\it not} to belong to model A.\cite{lidmar}

In light of this situation, it is important 
to test the conventional theory by performing a scaling analysis, in which
the dynamical exponent, $z$, and the scaling functions are {\it a priori}
unspecified.  $z$ is then determined by collapsing different data sets onto
a single curve.  Although conventional KTB dynamical theories do not
support such a general approach, a scaling ansatz has
recently been proposed by Fisher, Fisher, and Huse\cite{ffh} (FFH) for
superconducting critical phenomena.  This successful
ansatz has been applied to a wide variety of systems and transitions, 
including simulations of KTB dynamics.\cite{leeteitel}  We expect
that dynamical scaling will be appropriate in the same regimes where the
static KTB theory is applicable.

In the FFH theory, for 2D superconductors, 
the $I$-$V$ curves should scale as
\begin{equation}
\label{ffheq}
V=I\xi^{-z} \chi_\pm(I\xi/T), 
\end{equation}
where $\chi_{+ (-)} (x)$ is the scaling function for temperatures 
above (below) $T_{KTB}$.  
The two important asymptotic behaviors of $\chi(x)$ are
$\lim_{x\rightarrow 0} \chi_+ (x)= \mbox{const.}$ (ohmic limit), 
and $\lim_{x\rightarrow \infty}\chi_\pm (x)\propto x^z$ 
(critical isotherm). The universal jump appears as the difference in 
(log-log) slopes 
between the two asymptotic limits of $\chi_+(x)$, as we will discuss 
below. 

It is convenient to rewrite the Eq.~(\ref{ffheq}) as  
\begin{equation}
\frac{I}{T}\Biggl(\frac{I}{V}\Biggr)^{1/z}=\varepsilon_\pm(I\xi/T)
\label{FFHsc}
\end{equation}
where $\varepsilon_\pm (x)\equiv x/\chi_\pm^{1/z} (x)$. The advantage of 
Eq.~(\ref{FFHsc}) over Eq.~(\ref{ffheq}) is that one can better judge 
the scaling, because only the $x$-scale is 
stretched in Eq.~(\ref{FFHsc}). In Eq.~(\ref{ffheq}), both the $x$-scale and
$y$-scale 
are stretched making it harder to judge a collapse of the scaled data. 
(Compare Fig.~4 of Ref.~\onlinecite{harris91} with 
Figs.~\ref{1scsc}-\ref{sfsc} here.)

\subsubsection{Connections between the two approaches}
\label{sec:approaches3}

While not apparent at first glance, the connections between the dynamic 
scaling approach and the conventional approach become clear when one 
considers the following. First, one must keep in mind that 
the conventional approach, as described in Section \ref{sec:approaches1},
is valid only in the limit $I\rightarrow 0$ while the dynamic scaling approach
is valid for finite currents. Secondly, the relationship of one to the other 
should be considered for an arbitrary value of $z$. [For the generalization
of the conventional theory to an arbitrary value of $z$, see the discussion 
around Eqs.~(\ref{Rnf}) and (\ref{RT})]. 

Taking these considerations into account and looking at the asymptotic 
limit $I\rightarrow 0$ of the dynamic scaling function, one finds that
the two approaches are indeed compatible:  
(i) both theories predict that
the critical isotherm ($T=T_{KTB}$) should be a power-law
$V\propto I^{z+1}$ [i.e., $\alpha(T_{KTB})=z+1$], 
(ii) for $T<T_{KTB}$, both theories agree that the voltage 
remains a power-law of the current
(iii) for $T>T_{KTB}$, both theories give $\alpha(T)=1$, with $R(T)$
defined as
in Eq.~(\ref{RT}). 

\section{Scaling Results}
\label{sec:results}
In this section we will apply the scaling theory [Eq.~(\ref{FFHsc})] of 
Fisher, Fisher, and Huse\cite{ffh} to transport data from superconductors, 
Josephson Junction arrays and superfluids. The universality of the scaled data
will then be checked. Preliminary results on both high-temperature
superconductors
(HTSC's) and low-temperature (conventional) superconductors (LTSC's) were 
presented in Ref.~\onlinecite{ammirata98} where it was concluded that
$z\simeq 
5.7\pm 0.3.$

A preliminary estimate for the value of $z$ 
indicates that $z\gg 2$ for these systems. The critical
$I$-$V$ isotherm ($T=T_{KTB}$) is easily identified on a log-log plot 
by the fact that it is straight (i.e., $\alpha (T_{KTB})$
is independent of current). The slope of this isotherm
[$\alpha (T_{KTB})=z+1$] gives an estimate for $z$.  A visual check for 
this condition on the 
$I$-$V$ data of Repaci {\it et al.,}\cite{lobb96} Vadlamannati {\it 
et al.,}\cite{rutgers91} Matsuda {\it et al.,}\cite{matsuda92}, or any 
of the others clearly shows that, indeed, $z\gg 2$. With this initial 
evidence, we move on to a more rigorous scaling analysis of the data.

To perform the scaling analysis, $\xi_\pm (T)$ in Eq.~(\ref{FFHsc}) 
must be specified. This can be done for superconductors and Josephson 
junction array by exploiting the ohmic limit of Eq.~(\ref{FFHsc}): 
$R(T)\propto\xi_+^{-z}$. (For superfluids, the thermal 
conductance $K$ is used in  place of the resistance: $K\propto\xi_+^{z}$.) 
For $\xi_-$, we will assume that the vortex correlation length is symmetric 
about the transition (modulo some pre-factor) in this section. We will
explore 
the validity of this assumption in Sec.~\ref{sec:correlation}. 

Note that in the dynamic scaling theory, there are no 
requirements for the temperature dependence of $\xi_\pm (T)$. 
In this work, we will assume that the KTB form,
$\xi (T)\propto \exp[\sqrt{b/(T/T_{KTB}-1)}]$, provides the most 
efficient parameterization of the correlation length. It is through 
this assumption that the explicit connection with KTB theory is made. 
Any other temperature dependence for $R(T)$ could be used to 
check the scaling collapse, which would leave the type of transition more 
ambiguous. 

We determine $R(T)$ in two ways where possible. The first method is to 
extract it from the ohmic part of the $I$-$V$ curves and the second method is
to digitize the $R(T)$ data. The two results are then compared to 
one another. In the case of discrepancies, 
the $R(T)$ determined from the $I$-$V$ is used since thermal equilibrium 
is more likely in that case. Another advantage of using $R(T)$ 
determined from the $I$-$V$ is that one is assured that the 
$R(T)$ is ohmic. The disadvantage of course is that fewer data 
points are available for this $R(T)$. It should also be noted that we 
fit $R(T)$ only over the temperature range over which we have 
$I$-$V$ isotherms.

The fitting parameters for Eq.~(\ref{FFHsc}) are $z$ 
(universal), $T_{KTB}$ and $b$ (non-universal).  Three 
requirements must be fulfilled self-consistently in 
our scaling procedure:
(i) $V\propto I^{z+1}$, along the critical isotherm $T=T_{KTB}$; 
(ii) $R(T)\propto \xi^{-z}$, in the high temperature range; and 
(iii) scaling collapse of the $I$-$V$ isotherms, according to 
Eq.~(\ref{FFHsc}). Condition (i), which says that the $I$-$V$ 
curves are straight on a log-log scale at $T_{KTB}$, is used 
first to estimate a value of $T_{KTB}$ and $z$. That 
value of $T_{KTB}$ is then used in (ii) to fit the ohmic 
resistance data to obtain an expression for $\xi_\pm(T)$. 
Finally, condition (iii) is checked. Because there
may be a couple isotherms that appear to be straight on the log-log 
scale, this process 
is repeated, in the manner of Shaw {\it et al.}\cite{shaw96}, for the 
acceptable range of $T_{KTB}$'s to satisfy all three conditions.

In some of the data sets that we examined, the $I$-$V$ curves crossed 
over to an ohmic region at large $I$.\cite{herbert98} This behavior 
is not due to the 
critical behavior of the vortices but rather to a breakdown of the 
underlying superfluid. For that reason, we have omitted such data 
from our analysis.

Note that in the following, we have displayed all the scaling results 
(except for a few noted exceptions) without first making a judgment
of the quality of the data. As a result, the quality of the scaling
also varies. Nonetheless, we stress that each scaling result displayed 
here has been optimized for the best collapse and not for a value of $z$
in agreement with the other samples. This makes the result that all of the 
collapses occur for roughly the same value of $z$ all the more striking. 
Furthermore, we believe that there is a strong correlation between the 
quality of the raw $I$-$V$ data and the scaled data. But we leave 
this for the reader to judge.

\subsection{Superconducting films}
\label{sec:scscaling}
\begin{figure}
\centerline{
\epsfig{file=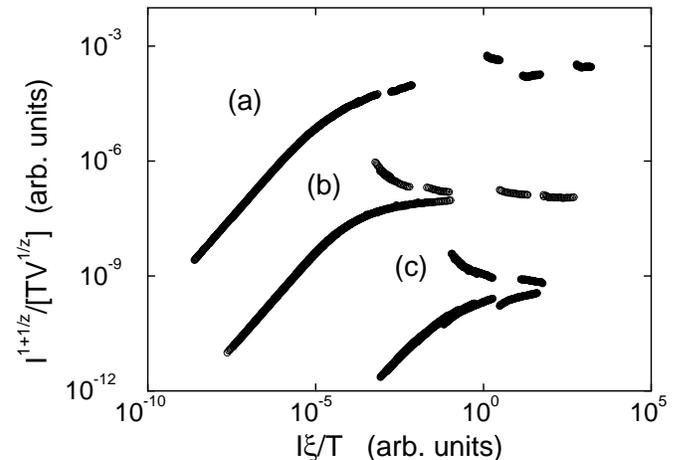,angle=-90,width=3.4in}}
\caption{The $I$-$V$ curves from thin superconducting
films scaled with 
Eq.~(\protect\ref{FFHsc}): (a) 24 $\AA$ thick YBCO layers in a 
multi-layer structure from  
S.~Vadlamannati {\it et al.};\protect\cite{rutgers91} (b) 24 $\AA$ 
thick YBCO mono-layer from Repaci {\it et al.};\protect\cite{lobb96} 
and (c) 100$\AA$ thick Hg/Xe film from Kadin {\it et al.}\protect\cite{goldman83} 
(Data sets (b) and (c) have been shifted arbitrarily.) The lower branch of these 
plots correspond to $T>T_{KTB}$. The limits of this branch are ohmic in the weak 
current limit to $V\propto I^{z+1}$ in the high current limit. It is these limits 
that represent the jump in the exponent $\alpha$.}
\label{1scsc}
\end{figure}

Fig.~\ref{1scsc} shows the scaling of three separate data 
sets to Eq.~(\ref{FFHsc}), (previously reported in 
Ref.~\onlinecite{ammirata98} but plotted on a different scale here.)
The first data set\cite{rutgers91} (marked $a$ in 
Fig.~\ref{1scsc}) covers a temperature range 
[30.06 K:46.09 K] and is from a YBCO/PrBa$_2$Cu$_3$O$_{7-\delta}$ 
multi-layer in which the YBCO layers have a thickness 
of $24~\AA$ and are reported to be nearly electrically isolated from one 
another by PrBa$_2$Cu$_3$O$_{7-\delta}$ (PBCO) barrier 
layers.  The scaling procedure leads to the results 
$T_{KTB}=32.0$~K, $b =14.0$, and $z=5.6\pm0.3$ where 
the resistance was fit over the range [43.5 K:47.0 K].
(See Figure \ref{resfit}.) Curve $b$ in Fig.~\ref{1scsc} is the 
scaled $I$-$V$ data from Ref.~\onlinecite{lobb96} taken 
on a $12~\AA$ thick YBCO mono-layer and includes isotherms ranging 
in temperatures from 10K to 40K. The resistance was fit over 
the range 25K to 34.5K yielding, along with the other two 
criteria, the following parameter values: $T_{KTB}=17.6$~K, 
$b=7.79$, and $z=5.9\pm0.3$. The scaled data set denoted by 
$c$ in Fig.~\ref{1scsc} and covering the temperature range 
[2.6 K:3.4 K] corresponds to a conventional, $100~\AA$ thick, 
superconducting sample\cite{goldman83} (Hg-Xe alloy). The parameters which
led to this collapse are $T_{KTB}=3.04$~K, $b=3.44$, and $z=5.6\pm0.3$. 
The resistance was fit over the temperature range [3.3 K:3.4 K]. 
While this collapse is not as complete as those of the others in this
figure, we emphasize that the best collapse was obtained for the 
reported value of $z$. 

A few features of the scaled data in Fig.~\ref{1scsc} should be 
pointed out. First, the upper branch corresponds to temperatures 
below the transition temperature and the lower branch to temperatures 
above the transition temperature. Secondly, an ohmic $I$-$V$ relationship
here is represented by a slope 1 on the log-log scale. One can see in 
each of the three curves in this figure (and the scaled data in the 
following figures,) the lower ($T>T_{KTB}$) branch is ohmic at low 
values of the scaling variable $x$ (typically low currents) and curves over 
and approaches a horizontal line as $x$ 
is increased. The horizontal line corresponds to the $I$-$V$ relation, 
$V\propto I^{z+1}$. 

\begin{figure}
\centerline{
\epsfig{file=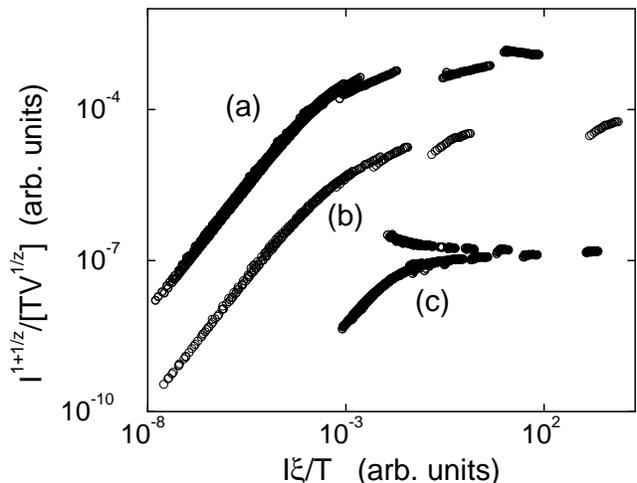,angle=-90,width=3.4in}}
\caption{The scaled $I$-$V$ curves of (a) 
Ref.~\protect\onlinecite{matsuda92} on a YBCO mono-layer; (b) 
Ref.~\protect\onlinecite{garland87} on a In/InO 150 $\AA$ thick composite film;  
and (c) Ref.~\protect\onlinecite{ammirata98} from a 1000 $\AA$ thick
BSCCO crystal. (Data set (b) and (c) has been shifted arbitrarily.) The collapse 
of curves (a) and (b) in this figure are not as complete as that of curve (c)
or curves (a) and (b) of Figure \protect\ref{1scsc}. Yet, the collapse could
not be improved using other values of $z$.}
\label{2scsc}
\end{figure}

For this paper, we have extended this analysis to two more 2D 
superconducting films. The collapse is shown in Fig.~\ref{2scsc}. 
The curve marked $a$ in that figure is from a mono-layer of YBCO 
sandwiched between two PBCO layers of different thicknesses\cite{matsuda92} 
and covers isotherms ranging from 16.32K to 41.31K. For the parameters, 
we find $T_{KTB}=18.3$~K, $b=31.04$, and $z=5.3\pm0.5$. The value of 
$T_{KTB}$ is similar to that of the YBCO mono-layer of 
Ref.~\onlinecite{lobb96} but the value of $b$ is nearly 
4 times larger. Curve $b$ in Fig.~\ref{2scsc} is from 
$I$-$V$ data on 150$\AA$ thick In/InO composite film from 
Ref.~\onlinecite{garland87}. The $I$-$V$ isotherms cover a 
temperature range [3.010 K:3.182 K]. Here, the resistance 
data determined from the $I$-$V$ curves covered only a limited range
and did not match well the $R(T)$ data from Fig.~3 of that
paper. For these reasons, the scaling criteria (ii)  
was not fully met.  Nevertheless, the scaling collapse occurred 
for a value of $z$ near that of the other samples: 
$T_{KTB}=2.97$~K, $b=10.21$, and $z=5.2\pm0.5$.

We also applied this scaling analysis\cite{ammirata98} to a 
thicker ($1500 \pm 500~\AA$) Bi$_2$Sr$_2$CaCu$_2$O$_{8+\delta}$ 
crystal. For such a thick crystal of a layered material, one 
would expect a crossover to 3D behavior near the critical 
temperature and a failure of the 2D scaling and perhaps a 
breakdown in the scaling. We however did 
not see any breakdown in the scaling for this sample. Indeed, 
as shown in curve $c$ of Fig.~\ref{2scsc}, we found a good 
collapse of the data with the 2D scaling form 
with $z=5.6\pm0.3$, $T_{KTB}=78.87$~K, and $b=0.20$.
(See also Fig. 2 of Ref.~\onlinecite{ammirata98}.) A difference
between this scaled data and the others was found when we looked 
at universality.  
We examine this issue in Section \ref{sec:universe}
and discuss the reason why a thicker crystal may scale in the 
same way as the thinner samples in Section \ref{sec:3De}.

A scaling analysis of the Hebard and Fiory $I$-$V$ data on a In/InO 
film\cite{fiory83,hebard83} was not possible since the temperatures 
of their isotherms were not published.

\subsection{Josephson junction arrays}
\label{sec:jjascaling}

$I$-$V$ characteristics of Josephson junction arrays are expected 
to be similar to that of superconducting films.\cite{lobb83} 
In this section, we apply Eq.~(\ref{FFHsc}) to data from two 
JJA systems.\cite{vanderzant90,herbert97}  A primary difference 
for these systems is that their resistance is not described by 
Eq.~(\ref{RT}) because the temperature is renormalized and depends 
upon the temperature-dependent critical current $i_c(T)$. In the 
data that we analyze below, we did not have access to $i_c(T)$ and 
so we could not determine if $R(T)$ followed the KTB behavior for 
JJA's.  Nonetheless, following the discussion at the beginning of 
this Section (\ref{sec:results}), we will use the Eq.~(\ref{RT}) 
to check for a collapse of the data according to Eq.~(\ref{FFHsc}). 
As a result, the value of $b$ will not have the significance it had 
in Section \ref{sec:scscaling}. 

\begin{figure}
\centerline{
\epsfig{file=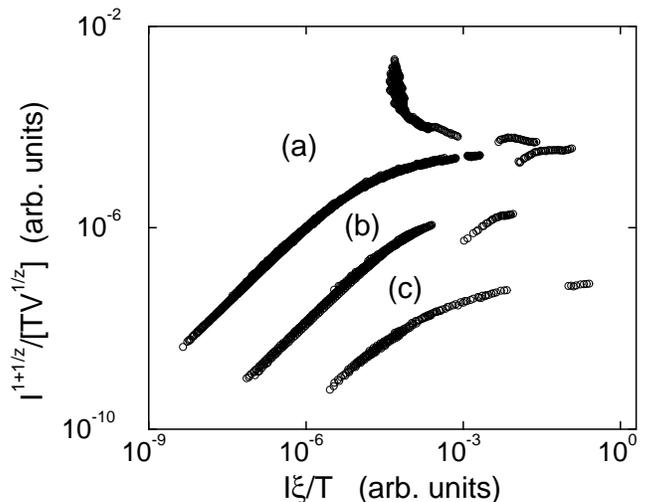,angle=-90,width=3.4in}}
\caption{The scaled $I$-$V$ curves from Josephson 
junction arrays: (a) YBCO/Ag weak links from Ref.~\protect\onlinecite{herbert97}; 
(b) Nb-Ag-Nb proximity-coupled junctions from Ref.~\protect\onlinecite{harris91}; 
and (c)  Nb/Nb arrays of Ref.~\protect\onlinecite{vanderzant90}.
(Data sets (b) and (c) has been shifted arbitrarily.)}
\label{JJAsc}
\end{figure}

Curve $a$ in Fig.~\ref{JJAsc} is scaled $I$-$V$ data from an high-temperature 
superconducting Josephson junction array (YBCO/Ag)\cite{herbert97} 
where the parameters used were $T_{KTB}=74.3$~K, $z=5.8\pm0.4$ and 
$b=0.72$. The resistance was fit over the temperature range: [78K:90 K]. 
One can see that the collapse is very good except for the 
isotherms furthest below $T_{KTB}$. This breakdown could be because
those isotherms are out of the critical region. An attempt to optimize 
the $T\ll T_{KTB}$ collapse by letting the correlation length be 
asymmetric in accordance with Eq.~(\ref{xi-}) was unsuccessful.

The scaled data denoted by curve $b$ in Fig.~\ref{JJAsc} is from 
a Nb-Ag-Nb proximity-coupled 
junction array.\cite{harris91,harris89} (We used the ``100\%'' data
from Ref.~\onlinecite{harris89}.) The parameters which produced the
best fit are $T_{KTB}=6.84$~K, $z=5.8\pm0.3$ and $b=0.32$ and the resistance
fit was over the temperature range: [7.3K:7.8 K]. The scaled I-V's
covered the range: [6.9K:7.8 K]. One can see that he 6.9K data set which
does not scale well for low currents. We believe that this is not a real
effect since all of that non-collapsing data has a voltage $V<10^{-9}V$ 
and does not parallel the behavior of the other data in that range. The data
do scale well for $V>10^{-9}V$. Harris {\it et al.}\cite{harris91} have
previously checked the dynamic scaling of this data and concluded 
that $z=2$. We believe that they would have reached the same result as found
here had they allowed $z$ to vary to optimize the scaling and had they used 
Eq.~(\ref{FFHsc}) instead of Eq.~(\ref{ffheq}). 

Curve $c$ in Fig.~\ref{JJAsc} is the scaled $I$-$V$ data 
from a Nb/Nb Josephson junction array.\cite{vanderzant90} 
The parameters used to optimize this were $\tau_{KTB}=0.51$~K, 
$z=5.5\pm0.5$ and $b=5.7$, where $\tau=k_BT/[(\hbar/2e)i_c(T)]$. 
A collapse of the scaled data 
could be obtained over a relatively large range of $z$ and 
$T_{KTB}$. We attribute this to the fact that the $I$-$V$'s 
isotherm only covered 2-3 decades of voltage and that there 
were no isotherms for temperatures below $T_{KTB}$. The resistance
was fit over the range $\tau=$ [0.8:1.3].

In a conventional analysis, Abraham {\it et al.}\cite{abraham82} 
reported a jump in 
the $I$-$V$ exponent $\alpha$ of 3 to 1 for PbBi/Cu arrays, a 
result which implies $z=2$. As far as we could tell, that data 
was not published and so was not available for our analysis. We 
tried to apply Eq.~(\ref{FFHsc}) to sample 6-18-81 of 
Ref.~\onlinecite{abraham83} but no definitive conclusions were 
reached due to the fact that the $I$-$V$ isotherms only 
covered 2-3 decades of voltage and a limited temperature range.

\subsection{Superfluid $^4$He films}
\label{sec:sfscaling}

For superfluid $^4$He films, the analog of electrical conductance and 
$I$-$V$ curves are thermal conductance and $\dot Q$-$\Delta T$ curves, 
where $\dot Q$ is power through the film and $\Delta T$ is the 
temperature gradient across the film. These measurements are done by
injecting heat at one end of a thin superfluid film adsorbed on a surface (e.g. mylar).
Because of dissipation from vortex pairs, there is a temperature gradient across the
film that is measured and is the analog of the voltage in the superconducting
measurements. In 
reality, $\dot Q$ is not the heat through the film but rather represents
$^4$He mass flow from the cool end of the film to the warm end, which occurs to 
replenish the $^4$He which evaporates from the warmer end at a quicker 
rate than from the other end. This is thoroughly discussed in
Ref.~\onlinecite{maps83}.

As we shall shortly see, these results do show that $z\simeq 5.6$. However,
we are careful to point out that the results we are about to 
present can only be said to be consistent with such
a value of $z$ but cannot be taken to be evidence that 
$z\simeq5.6$ for two reasons. 
First, there is no reliable thermal conductance $K$ data 
in the ``ohmic'' limit (i.e., $T>T_{KTB}$ zero power $\dot Q$ limit.) 
Furthermore, one has to account for the conductance of the 
gas $K_g$ surrounding the film which can only be approximated
to within a factor of two. These two points allow us more freedom 
to obtain the optimal scaling. Nonetheless, the best scaling 
does yield a $z$ which is similar to that of superconductors 
and Josephson junction arrays.

\begin{figure}
\centerline{
\epsfig{file=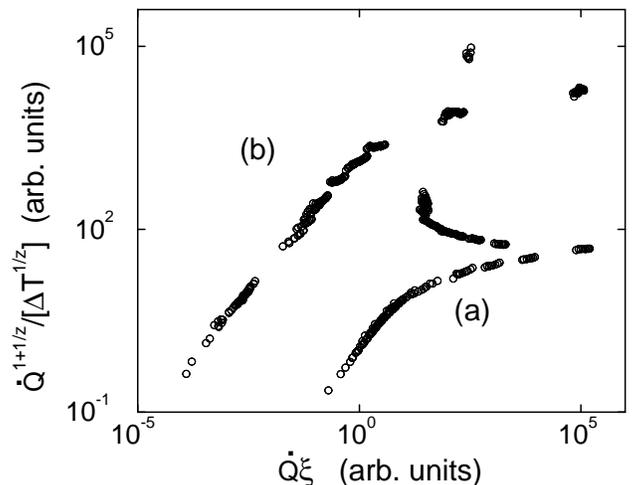,angle=-90,width=3.4in}}
\caption{The scaled $\dot Q$-$\Delta T$ curves from superfluid 
$^4$Helium films of (a) Ref.~\protect\onlinecite{maps83}; and (b)  
Ref.~\protect\onlinecite{dionne87}. (Data set (b) has been shifted arbitrarily.)}
\label{sfsc}
\end{figure}

The collapsed data in Fig.~\ref{sfsc} marked curve $a$ is from 
Ref.~\onlinecite{maps83}. Instead of varying the temperature, 
those authors varied film thickness $d$ which in turn varies 
the transition temperature. The independent variable then is $d$ 
and not $T$ and so the correlation length depends upon $|d-d_c|$ 
instead of $|T-T_{KTB}|$ where $d_c$ is the thickness whose $T_{KTB}$ 
corresponds to the temperature of the experiment. (See Eq.~(17) of 
Ref.~\onlinecite{maps83}.)  By adjusting $d$, they were able to obtain 
$\dot Q$-$\Delta T$ data for both above and below the transition 
temperature. The parameters that we obtained were $d_c=5.4$, 
$z=5.4\pm0.4$, and $K_g=8.0\times10^{-4}$ W/K. We assumed that 
$K_g$ was a constant over the parameter space that the $\dot Q$-$\Delta T$ 
curves covered and estimated it based on the small $d$ behavior of 
$K$. This placed some limits on this quantity but we were still 
able to vary it by a factor of two. We found that the scaling 
collapse was relatively insensitive 
to the value of $K_g$ because most of the power was flowing through 
the film. As one can see, the scaling starts to break down for 
thicknesses ($d\sim9$ layers) much larger than $d_c$. 

Curve $b$ in Fig.~\ref{sfsc} is from Ref.~\onlinecite{dionne87}. It
contains more data corresponding to $T>T_{KTB}$ but
doesn't cover as wide a range of thicknesses overall as the data of Maps 
{\it et al.}\cite{maps83} The collapse was found to be consistent 
with the results we have presented in this paper: $d_c=5.2$, 
$z=5.6\pm0.5$, and $K_g=5.0\times10^{-4}$ W/K. Like the JJA data of
Ref.~\onlinecite{vanderzant90}, the error bars here are large because the 
data extends only over a couple orders of magnitude.

\subsection{Universality}
\label{sec:universe}
One of the fundamental properties of critical behavior and scaling is 
universality: the idea that the same function 
$\varepsilon_\pm(x)$ in Eq.~(\ref{FFHsc}) and the same 
value of $z$ describes 
all of the data independent of the system or material. 
For the $T>T_{KTB}$ branch, we found universality for 
nearly all of our scaled data. For the $T<T_{KTB}$ branch, 
on the other hand, the same function $\varepsilon_\pm(x)$ 
described the data sets of the thinnest 2D samples but not that 
of the layered materials\cite{ammirata98,rutgers91} or the superfluid 
Helium.\cite{maps83} We explore these issues in this section.

\begin{figure}
\centerline{
\epsfig{file=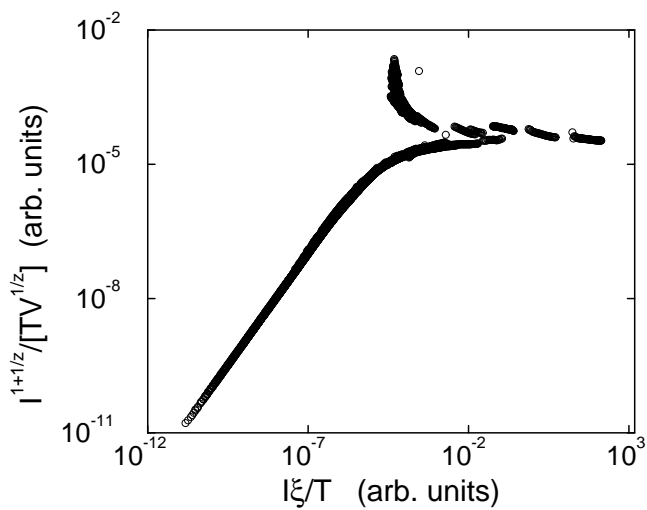,angle=-90,width=3.4in}}
\caption{
The JJA data of
Fig.~\protect\ref{JJAsc}a\protect\cite{herbert97}, the YBCO SC 
data of Fig.~\protect\ref{1scsc}b\protect\cite{lobb96}, and the Hg/Xe SC data 
of Fig.~\protect\ref{1scsc}c\protect\cite{goldman83} plotted together. The
scaled data 
sets have been shifted to show universality of the scaling function.}
\label{universalityfig1}
\end{figure}

In Figure \ref{universalityfig1}, the HTSC data from the 
YBCO mono-layer,\cite{lobb96} the LTSC data
from the Hg/Xe thin film,\cite{goldman83} and the YBCO 
JJA data\cite{herbert97} are plotted together. Because 
the scaling functions are dimensionless, we multiply the 
$x$ and $y$ axes by non-universal constants which enter 
the scaling theory to account for the sample dependence. 
In order to test universality, all the data sets must be
scaled with the same value of $z$. So we have adjusted
each data set within the error bars of the parameters
so that each has $z=5.6$. 
As one can see, the collapse is very good, and bolsters 
the evidence that this data scales and that $z\simeq 5.6\pm 0.3$. 
The agreement of the Repaci {\it et al.}\cite{lobb96} 
and the Herbert {\it et al.}\cite{herbert97} data is 
particularly impressive since both data sets have 
extensive $T<T_{KTB}$ branches and because they come 
from two different systems: JJA's and SC's.

\begin{figure}
\centerline{
\epsfig{file=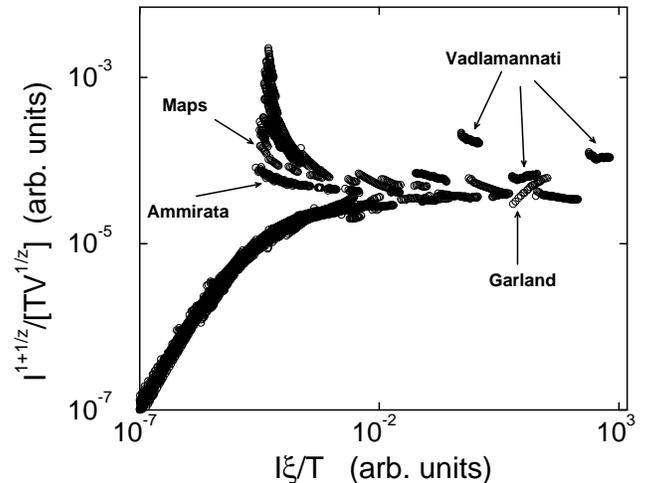,angle=-90,width=3.4in}}
\caption{The data from all of the data sets plotted together. 
The $T>T_{KTB}$ data collapse very well suggesting universality of the 
scaling function for that temperature regime. The $T<T_{KTB}$ data also scale 
well if three data sets, belonging to layered materials (labeled 
Ammirata\protect\cite{ammirata98} and Vadlamannati\protect\cite{rutgers91}) 
and superfluid Helium (labeled Maps\protect\cite{maps83}), are neglected. 
This is discussed in the text. The Garland and Lee data\protect\cite{garland87} 
scaled well except for the anomalous isotherm just above $T_{KTB}$, labeled 
Garland.\protect\cite{garland87}}
\label{univall}
\end{figure}

In Figure \ref{univall}, we plot all of the SC, 
JJA, and SF data that we have presented here in a single 
plot to test universality. As in the previous figure,
we adjusted each data set within the error bars so that 
$z=5.6$. We also removed some of the individual scaled isotherms
that did not scale well with the other scaled isotherms from
the same sample. This includes two of the isotherms from
the Matsuda {\it et al.}\cite{matsuda92} data (the second and
third from the right in data set (a) of Fig.~\ref{2scsc}), 
parts of two of the data sets from data set (b) of Fig.~\ref{2scsc} 
(second and third from the right), and part of one
data set from Harris {\it et al.}\cite{harris91} (the rightmost set
labeled (b) in Fig.~\ref{JJAsc}). One can see clearly that the 
$T>T_{KTB}$ data scales well and strongly suggests 
universality for this temperature regime. The Garland and 
Lee\cite{garland87} data is the weakest of these since it's lowest
temperature isotherm does not lie on the other scaled data as
we have pointed out in this figure.

Universality was not found for all of the $T<T_{KTB}$ data. Besides 
the data plotted in Figure~\ref{universalityfig1}, there 
are only a few data sets that have significant $T<T_{KTB}$ 
branches: the HTSC BSCCO thin crystal of 
Ref.~\onlinecite{ammirata98}, the $^4$Helium data 
of Ref.~\onlinecite{maps83}, and the YBCO multi-layer 
data of Ref.~\onlinecite{rutgers91}. (We have labeled each of these
data sets in Fig.~\ref{univall}.) The low temperature 
branches of these data sets fail to collapse with the 
other data. It is likely that the BSCCO and YBCO multi-layer 
data fail to collapse in this temperature range due to their 
layered structure. We will discuss this point in Section \ref{sec:3De}. 
As to why the $T<T_{KTB}$ $^4$Helium scaled data
do not collapse with the other data, a possible explanation is 
that the approximate determination of $\xi(T)$ was inaccurate. (Recall 
the lack of ``ohmic" conductivity data for the superfluid measurements.) 
We recommend further studies of such data to more accurately determine 
$\xi(T)$ and to check the universality. In short, we cannot conclude 
whether the lack of 
universality of the scaling function for $T<T_{KTB}$ indicates a
breakdown of the dynamic scaling results, or if it is due to the data 
or systems considered. We believe it is the latter.

\section{Discussion}
\label{sec:discussion}
The main result of Section \ref{sec:results} is that $z\simeq 5.6$ 
for superconductors, superfluids, and Josephson junction arrays. 
The large value of $z$ is nearly three times the expected 
value for these systems. Indeed there are many reports for 
superconductors and Josephson junction arrays that the $I$-$V$ 
exponent $\alpha$ jumps from 3 to 1 at the transition 
temperature, a result that is consistent with $z=2$. In 
this section, we explore the discrepancy between the approaches by 
discussing the validity of each. 

\subsection{Dynamic Scaling}
\label{sec:validityDS}

Dynamic scaling is a powerful technique which has proved particularly
useful in

investigating the nature of the $H$-$T$ phase diagram in the high-temperature 
superconductors. Yet, one must be careful with the results obtained with a
scaling 
analysis because this technique is not without its weaknesses. For example, in
Section
\ref{sec:universe}, universality of the scaled data was examined and, while
universality was found for the $T>T_{KTB}$ data and some of the $T<T_{KTB}$
data, it did not hold for all of the latter. In this Section, 
we will examine other aspects of the dynamic scaling 
analysis presented here and its validity.

\subsubsection{Vortex correlation length}
\label{sec:correlation}
There are various aspects of assumptions that we made regarding
the vortex correlation length that may detract from the dynamic 
scaling analysis. Here we make considerations on 
its symmetry, the value of the non-universal 
constant $b$, and the expected range of the validity of the 
temperature dependence.

Throughout Section \ref{sec:results} it was assumed that the 
vortex correlation length was symmetric about the transition 
temperature even though Ambegaokar {\it et al.}\cite{ahns80} predict
that it should be asymmetric. [See Eq.~(\ref{xi-}).] 
Surprisingly, where there was data for both 
above and below the transition temperature, the scaling did 
not seem to suffer from this assumption. This is 
particularly so for the data on the superconductors. 
One notices that for the data of Refs.~\onlinecite{goldman83}, 
\onlinecite{lobb96}, and \onlinecite{ammirata98}, 
the scaling works well for both branches. (See Figures \ref{1scsc} and
\ref{2scsc}.) 

When we allowed the correlation length to be asymmetric by allowing for
different
values of $b$ below $T_{KTB}$ ($b_-$), and above $T_{KTB}$ ($b_+$),
significantly better scaling could not be achieved, possibly 
indicating that $\xi$ is symmetric, in accordance with
the scaling of the numerical results of Lee and Teitel.\cite{leeteitel} 
There were of course a few exceptions,\cite{herbert97,maps83} 
but in neither case was 
prediction $b_+=2\pi b_-$\cite{ahns80} verified. In 
fact, in these two cases, $b_-$ tended to be larger 
than $b_+$. For the superfluid $^4$He data of 
Ref.~\onlinecite{maps83}, we found $b_-\simeq3b_+$. 
We do not view this result as conclusive because of the afore-mentioned
problems with determining $\xi(T)$ for helium and we suggest 
further study of this topic. 

In our fitting of the resistance $R(T)$ to determine $\xi_\pm(T)$, 
we were sometimes able to fit the resistance to the Kosterlitz 
form over an extended temperature range. A notable example is the 
YBCO mono-layer data\cite{lobb96} where we fit the resistance from 
25 K to 35 K. The upper limit of this fit is twice the $T_{KTB}$ 
which is remarkable since true KTB critical behavior is expected 
to be valid over a very narrow temperature range.\cite{minnhagen87,friesen96}
This result could suggest that the critical region is larger than 
expected, or more likely that KTB-like behavior remains 
valid outside the critical region.\cite{minnhagen87} Indeed, Minnhagen
and Olsson\cite{minnhagen92} indicate that Eq.~(\ref{xi+}) 
is a useful phenomenological form over wider temperature regimes as long
as $b$ is taken to be a phenomenological parameter.\cite{leeteitel} We 
adopt this interpretation.

In the literature, it is commonly stated that $b$ is material 
dependent but that $b=O(1)$.\cite{k74} In our results for 
superconductors, we found $b$ to roughly $O(10)$ but as small as 3.44 and
0.2, 
thereby varying by an order of magnitude or two. We also 
found that its value could vary within materials. For example, the value 
of $b$ for the 2 YBCO mono-layer systems examined in Section
\ref{sec:scscaling} 
varied by a factor of 4,\cite{lobb96,matsuda92} which is concerning.
However, we do not think it is troubling that
the $b$ varied from material to material by an order of magnitude. The
systems 
considered here are diverse from one another. For example, the electron
density can vary significantly from the conventional superconductors to the HTSC's. 
Another interpretation is that the value of $b$ is a phenomenological one and
not equivalent to its true asymptotic critical
value.\cite{leeteitel,minnhagen92}

\subsubsection{Universal jump: 6.6 to 1}
\label{sec:unijump}
If one is to believe that $z\simeq 5.6$ in these materials, then 
one expects to see a jump in the $I$-$V$ exponent $\alpha$ of 6.6 
to 1 in the $I\rightarrow 0$ limit. Yet evidence of this is not observed 
in any of the samples. We believe this is because previous measurements 
have not gone to weak enough currents to observe this behavior. For 
example, based on the scaling curve of the YBCO mono-layer 
data\cite{lobb96} in Fig.~\ref{1scsc}b where $T_{KTB}=17.6$K, 
one can see that the scaling curve is ohmic for $x\lesssim 10^{-5}$.
($x$ is the scaling variable.) This 
means that the 18 K isotherm would not become ohmic until 
the current $I\lesssim 10^{-11}$A. Clearly, voltage 
sensitivity is far from detecting that crossover. [In fact, 
that isotherm would become first ohmic due to finite size effects 
at a much higher current (Sec.~\ref{sec:fse}).] It is a similar
situation for the Herbert {\it et al.} data\cite{herbert97} where 
the scaling curve is ohmic for $x\lesssim 10^{-6}$. The $74.722$~K 
isotherm, the first isotherm above $T_{KTB}=74.3$~K, would not become
ohmic until the current $I\lesssim10^{-8}$A. This is several orders of
magnitude smaller than the minimum current for that isotherm. 

The ``jump" is evident in the scaled data only in the following way. 
For the $T>T_{KTB}$ data, the scaled data goes from ohmic behavior 
($\alpha=1$) for small values of the scaling variable to 
$\alpha=6.6$ at large value of the scaling variable. In this way,
the ``jump" is only manifest as a crossover from the small $x$ 
behavior to the large $x$ behavior of the $T>T_{KTB}$ branch of
the scaled data.

In Section \ref{sec:convapp}, we will use a conventional approach to
show that the behavior of $\alpha$ for the YBCO mono-layer 
data\cite{lobb96} is consistent with the value of $z$ that we found
for that sample: $z\sim 5.9$.

\subsubsection{Three dimensional effects}
\label{sec:3De}

We have examined two samples which could be viewed as layered. (The first 
layered sample that we examined was the YBCO/PBCO multi-layer system of 
Ref.~\onlinecite{rutgers91} where the two unit-cell thick YBCO layers were 
believed to be electrically isolated. The second sample was the 1000$\AA$ 
thick BSCCO crystal.\cite{ammirata98}) In layered superconductors, 
three dimensional (3D) behavior is expected in a small region
of the current-temperature phase diagram near $T_{KTB}$. One could then
ask why such samples would scale in the same way as the much thinner samples
and why 3D effects are not manifest. 

To discuss the 3D effects, two new lengths are introduced. 
The first is related to the energy of the
vortex pair. This length is the Josephson length $\lambda_J=\gamma s$, 
which incorporates the effects due to Josephson coupling between the layers. 
($\gamma$ is the anisotropy ratio for layered superconductors and is 
equivalent to the ratio of the coherence length in the $ab$ planes to 
the coherence length in the $c$ direction: $\gamma \equiv \xi_{ab} /\xi_c$. 
$s$ is the distance between layers.) For separations less than $\lambda_J$, 
the vortices interact with the 2D (logarithmic) potential. 
For larger separations however, the potential becomes linear due to 
the Josephson coupling between the layers.\cite{cataudella90} (The 
Josephson coupling also introduces a term to the interaction energy 
for lengths less than $\lambda_J$, but that term is very small compared 
to the 2D logarithmic one.)

The second length is the $c$-axis vortex correlation length $\xi_c^v$ (to be 
distinguished from the $c$-axis superconducting coherence length) and 
characterizes how far along the $z$ direction the vortices are 
correlated. This length can also be viewed in terms of the length-scale 
dependent layer decoupling length $\ell_{3D/2D}=\gamma\xi_c^v$ 
(defined for $T>T_c$) since it determines
the extent over which the 3D effects are important in the in-plane 
direction.\cite{pierson} Because of vortex screening, the Josephson
interaction
is screened out beyond lengths $\ell_{3D/2D}$, making the interaction 2D at 
large separations for $T>T_c$. $\ell_{3D/2D}$ becomes small quickly above the
transition 
temperature.\cite{pierson}

Because of these two competing lengths, 3D behavior is expected only over a
small
range of the $I$-$T$ phase space. On the temperature axis, this region is 
limited by $s<\xi_c^v(T)<D$, where $D$ is the thickness of 
the sample. On the $I$ axis, the 3D region is limited to intermediate 
currents: $\lambda_J<r_c<\ell_{3D/2D}$. Above $T_{KTB}$, the renormalized
$\gamma$
(and hence $\lambda_J$) grows quickly\cite{friesen} while $\ell_{3D/2D}$,
because of the temperature dependence of $\xi_c^v$, decreases rapidly, 
further limiting the 3D behavior. This leaves only a small window in which 3D 
near $T_{KTB}$ effects can be observed. 

Returning our attention to the layered samples that we examined here, 
it is plausible that 3D effects are being seen. After all,
it is primarily the layered superconductors of Refs.~\onlinecite{rutgers91}
and \onlinecite{ammirata98} that don't obey universality. For
the multi-layer sample\cite{rutgers91}, the $T>T_{KTB}$ scaled data fails to
collapse with the other curves (see Fig.~\ref{univall}) for the
isotherms nearest $T_{KTB}$. That this sample could have 3D behavior
is not in contradiction with the reports of those authors that their layers
are electrically isolated\cite{rutgers91} since magnetic coupling extends over
larger distances.   For the BSCCO sample of Ref.~\onlinecite{ammirata98} whose
$T<T_{KTB}$ data does not collapse with the others in Fig.~\ref{univall}, 
we have done the following calculation based on our above discussion 
to try to estimate where 3D effects should be seen.
Since $\gamma \simeq 160$ for BSCCO 2212, 3D effects should start to occur
for currents $\lesssim 1.2$~mA (where $r_c\gtrsim\lambda_J$) and 
persist up to the current associated with the minimum of 
$L_{fs}$ ($\sim 2.4$mA) or the decoupling length, very near 
$T_{KTB}$. Based on these rough estimates, which do not incorporate
renormalization effects, it seems unlikely that 3D effects could 
be observed in these samples for $T>T_{KTB}$. For the $T<T_{KTB}$ branch, 
it is more likely that the deviation from the universal curve is due to the
thickness of that crystal since it is well known that the 3D region is 
much larger below the transition temperature than 
above.\cite{pierson95,friesen95}

\subsection{Validity of ``conventional" results}
\label{sec:validityCA}

In the previous Section (\ref{sec:validityDS}), we have 
addressed the validity of the dynamic scaling, whose results 
indicate that $z\simeq 5.6$. Since this contradicts the 
evidence from the conventional approach that $z=2$, we now 
address those results. We do not claim that each paper is incorrect 
in their claims of $z=2$ but we do hope to convince the reader that 
the question of the value of $z$ is still an open one. 

The conventional results fall into roughly two 
classes, dc and ac. The dc measurements are the most common and 
include the determination of $\alpha(T)$ from the $I$-$V$ 
measurements. dc magneto-resistance measurements have also been
used but less frequently. The ac measurements include the torsion oscillator
$^4$He measurements of Bishop and Reppy\cite{bishop78,bishop80} 
and the ac inductance measurements of Fiory, Hebard, and 
Glaberson.\cite{fiory83} Clearly, we cannot address each paper 
that reports evidence for $z=2$ and so we will discuss them in 
broad terms using particular examples where appropriate. 

\subsubsection{$I$-$V$ and $R(T)$ dc Measurements}
\label{sec:alpha3to1}

Most of the papers that report evidence for a KTB transition 
or KTB behavior make their determinations based only on $\alpha (T)$ 
and $R(T)$ measurements. (There are a few notable exceptions to 
this that we discuss below.\cite{fiory83,garland87}) We suggest 
here that such an approach cannot be taken as evidence of KTB 
behavior and $z=2$ but only as being consistent with $z=2$
within the conventional approach. 

We begin with a brief description of this method. From the 
$I$-$V$ data, a value of $T_{KTB}$ is determined from the 
condition, $\alpha(T=T_{KTB})=3$, (which, of course,  
assumes $z=2$). It is then that this value of the
transition temperature is consistent with the $R(T)$ data and 
the Minnhagen\cite{minnhagen87} form of the resistance:
\begin{equation}
\label{RTmin}
R(T)=A \exp[-2\sqrt{b(T_{c0}-T_{KTB})/(T-T_{KTB})}],
\end{equation}
or the traditional form for $R(T)$ [Eq.~(\ref{Rnf2}).]
The mean field temperature $T_{c0}$ and constant $A$, in addition to $b$
amount to three fitting parameters. (If Eq.~(\ref{RT2}) is used, then there 
are only two fitting parameters.) A further check that is sometimes used
is to verify that $\alpha(T)$ decreases quickly above $T_{KTB}$.

There are several reasons for why this approach can be a misleading
check of $z=2$. The first and most important is that determining where
$\alpha(T)=3$ is difficult. It is well known that the predicted 
jump in the $I$-$V$ exponent $\alpha(T)$ exists only in the 
$I\rightarrow 0$ limit\cite{goldman83,herbert98,lobb96,shenoy94} 
and that, for temperatures above the transition temperature, 
the value of $\alpha(I,T)$ can vary quickly from a value of 
$1$ to a value of $z+1$ as a function of current. Hence,
one must be sure that the value of reported $\alpha(T)$ will not 
dip to a lower value at currents whose voltages are below the 
voltage sensitivity.

(Beyond the problem of probing the weak current limit, actually 
detecting the jump can be very difficult because, 
in the presence of finite size effects, small 
magnetic fields, and disorder, the jump 
gets rounded considerably, due to the many competing length 
scales. Note that most 
papers\cite{rutgers91,hebard83,epstein81,norton93} 
do not report a jump in the $I$-$V$ exponent.)

The second problem with this approach is that, most of the time, 
a misleading criterion is used to determine $\alpha(T)$. Because
it is seldom (if ever) the situation that one knows that the 
value of $\alpha(T)$ determined for a particular isotherm 
represents the weak current limit, one should examine
$\alpha(T)$ for a given length scale (or common current), as 
originally suggested by Kadin {\it et al.}\cite{goldman83} 
Here, one looks for a change in the behavior of this quantity 
near the transition temperature. This approach is usually 
not followed rigorously however. Instead of using a common 
current, investigators will determine $\alpha(T)$ for a common 
voltage range. This has the effect of biasing the results because, as 
one looks at higher temperature isotherms at common
voltages, one is looking at lower currents. This means that $\alpha(T)$
will have decreased not only due to temperature but also due to longer 
length scales. Therefore, any report rapid decreases of $\alpha(T)$ may
be an artifact of examining the isotherms at a constant voltage. 
(There are also cases where not even a common voltage range is used;
rather, the parts of the isotherms that deliver the desired rapid 
change in $\alpha(T)$ are studied.) 

Thirdly, once a value of $T_{KTB}$ is obtained for where $\alpha=3$, 
the number of fitting parameters in Eq.~(\ref{RTmin}) used to
verify the self consistency is large (three) and the temperature 
range over which one fits the $R(T)$ data is limited. So we contend that the 
conditions for checking $z=2$ in this approach are not stringent 
enough to be considered as evidence.

Finally, we note the inherent problem with this approach mentioned above that 
$z$ is not allowed to vary in order to optimize the analysis. 

As an example, in Ref.~\onlinecite{herbert97} a jump in the 
$I$-$V$ characteristics is reported for the YBCO/Ag weak 
link JJA. (See Fig.~2 inset in Ref.~\onlinecite{herbert97}a 
or Figs.~5-7 and 5-8 in Ref.~\onlinecite{herbert97}b.) The sharp jump in 
Ref.~\onlinecite{herbert97} was obtained by fitting the ohmic part 
of the curves at a temperature just above where the $I$-$V$ 
has $\alpha=3$. In this case, no fit of $R(T)$ was done to check
the value of $T_{KTB}$ for consistency. Also,
one cannot rule out that, if those 
authors had another decade or two of voltage sensitivity, 
isotherms at temperatures lower than their $T_{KTB}$ 
would also become ohmic at lower currents. Further, 
as is clear from the scaling analysis shown in 
Figs.~\ref{JJAsc}a and \ref{universalityfig1}, this data 
is consistent with $z=5.6$. As a second example of this,
we look at the YBCO mono-layer data of Ref.~\onlinecite{matsuda92}.
By inspecting Fig.~4 of the reference, it is clear that some
of the isotherms below their reported $T_{KTB}$ could manifest ohmic
behavior if more decades of voltage sensitivity were possible.
An inspection of the $I$-$V$ data of Ref.~\onlinecite{rutgers91} (Fig.~2)
yields a similar conclusion. It is unlikely that the reported
value of $\alpha=3$ at $T=T_{KTB}=40.1$ is the asymptotic value of that 
quantity at low currents. [The voltage sensitivity for that data also 
occurs at a larger value ($\sim 10^{-7}V$) than that of other measurements 
($\sim 10^{-9}V$).] 

To summarize, $I$-$V$ along with $R(T)$ measurements do 
provide self-consistent results for $\alpha(T=T_{KTB})=3$ (and
thus $z=2$), but cannot be taken as proof that $z=2$. This is
because the flexibility in determining $T_{KTB}$ and the 
three fitting parameters used in Eq.~(\ref{RTmin})
to fit the smooth, monotonic $R(T)$ data do not pose tight 
enough constraints.

\subsubsection{Kinetic inductance on SC's, magneto-resistance, 
and $^4$Helium torsion experiments}
\label{sec:inductance}

We now turn to other measurements used to study dynamics in
the SC, SF, or JJA systems.

As mentioned above, there are a few notable papers that 
went beyond measuring only $I$-$V$ and $R(T)$ curves. 
The most comprehensive study was done by Fiory, 
Hebard and Glaberson\cite{fiory83} who looked at kinetic 
inductance and magnetoristance in addition to the usual 
$I$-$V$ and $R(T)$ measurements on In/InO systems. The 
thoroughness of their approach is commendable.
While their results seem to be consistent with $z=2$, 
they are not without their inconsistencies and cannot necessarily 
be taken as definitive evidence for $z=2$.

Fiory {\it et al.}\cite{fiory83} were able to determine the vortex
interaction strength $q^2$ in 
two ways for a rectangular strip sample of In/InO. 
In their $ac$ impedance measurements used to determine the
kinetic inductance, the frequencies are small (160 Hz) and so they are
probing only statics. In that case the kinetic inductance 
does not depend upon the density of free vortices
$n_f$ but only the superfluid density. In that case, $q^2$ 
is directly determined. In their $I$-$V$ measurements, a dynamic
value of $q^2$ is determined. They find that the
values of $q^2$ from the two measurements agree over a 
limited temperature range ($\sim30$mK), which would be 
consistent with a value of $z=2$. (See Fig.~9 of 
Ref.~\onlinecite{fiory83}.) However, this result should be
viewed cautiously since the two measurements disagree for 
most of the region $T<T_{KTB}$. Further, the four-probe 
contact method that they used to measure the data  for the comparison of the 
values of $q^2$ is less accurate and less sensitive\cite{fiory83} 
than the two-coil contactless method that they used on
circular sample from the same film. Moreover, their measurements
of samples from the same film revealed variations in $T_c$ that suggest
that sample inhomogeneities may be significant.

To further test the value of $T_{KTB}$ for the In/InO films, 
Fiory {\it et al.}\cite{fiory83} measured the magneto-resistance $R(H)$. 
According to the theory of Minnhagen,\cite{minnhagen81} $R(H)$ should 
be linear in $H$ at $T_{KTB}$ but sub-linear above it and faster than 
linear below it. Their data at $T=1.782$K do show a region which 
is nearly linear ($R\propto H^{1.07}$) over roughly two decades. 
This temperature for crossover is roughly consistent with their 
$T_{KTB}=1.782$K determined from $\alpha(T=T_{KTB})=3$ criteria. 
As mentioned above, the roughly linear area is over only two decades
and the samples do have a degree of inhomogeneity to them. Further, it
is likely that surface barriers should be taken into account.\cite{fuchs98}

Garland and Lee\cite{garland87} have also used magneto-resistance data 
in addition to the $I$-$V$ and $R(T)$ measurements on In/InO films. 
Based on their Minnhagen criteria for $R(H)$ they determine 
$T_{KTB}=3.123$K. At this temperature, they find that $\alpha(T)$ 
drops from a value of roughly four to one. One will also notice that
their $I$-$V$ at $T_{KTB}$ is not a pure power law as required by the 
dynamic scaling and that their $T<T_{KTB}$ isotherms all have a positive 
curvature. They attribute this behavior to field-induced vortices. 
It is clear from Fig.~7 in their paper that the field plays a 
role for fields at least as low as 5 mG. Yet, in keeping with the 
discussion on finite size effects in Section \ref{sec:fse}, 
it is our view that the crossover 
to ohmic $I$-$V$ in their Fig.~5 should occur at a roughly common 
value of the current. This is because the magnetic length scale should not 
change significantly over the 100 mK that their $I$-$V$ isotherms 
cover below their claimed $T_{KTB}$. This is not observed in that data.
Choosing $T_{KTB}$ at a lower temperature
seems to be a better explanation, especially when one considers the 
good collapse of their data in the dynamic scaling analysis as seen in 
Figure \ref{2scsc}b. 

$^4$Helium torsion experiments\cite{bishop78,bishop80} 
were among the first evidence for the 
Kosterlitz-Thouless-Berezinskii transition.  
Nevertheless, while we agree that these measurement 
are indicative of KTB behavior (i.e., vortex 
pair unbinding), we do not believe that they indicate 
$z=2$. In the case of $^4$Helium torsion 
experiments\cite{bishop80}, Bishop and Reppy measured 
the period shift and $Q$ value of an Andronikashvili 
cell. The former is proportional to the real part of 
the dielectric constant $\epsilon$ and the latter to 
the imaginary part of that quantity. It is the latter 
that has the predominant dependence on the free vortex 
density $n_f$. [See their Eq.~(A2) or Eqs.~(3.17) and 
(3.18) in Ref.~\onlinecite{ahns80}.] They implicitly 
assume $z=2$ in writing Im$(\epsilon)\propto n_f$. We 
point out that their method cannot distinguish a value 
of $z$ but only the product $bz$. A value of $z$ other than 2 
would still make the values of their fitting parameters 
reasonable. The noise spectrum measurements of Shaw {\it 
et al.}\cite{shaw96} are another good example of a 
measurement that is not able to determine the value 
of $z$ but only the product $bz$.\cite{noteshaw} 
Hence these measurements cannot be taken as evidence for $z=2$.

\subsection{Comparison of Conventional Approach with Dynamic scaling}
\label{sec:comparison}

To further examine the validity of the finding $z\simeq5.6\pm0.3$ obtained 
from the Fisher-Fisher-Huse scaling, we will examine a data set using 
the conventional approach with an arbitrary value of $z$. 
We have chosen the YBCO mono-layer data of 
Ref.~\onlinecite{lobb96} for several reasons. First, their data 
covers by far the largest current, voltage, 
and temperature range of any data set in the literature. 
Secondly, those authors have pointed
out that there data does not satisfy the ``conventional" criteria for KTB
behavior:
the $I$-$V$ exponent does not vary rapidly from 3 to 1 near the transition 
temperature. Finally, they suggested finite size effects to explain the
lack of KTB behavior and performed a largely qualitative analysis.

In this section, we will use a quantitative analysis to show that 
finite size effects cannot account for the observed behavior in that 
data (Sec.~\ref{sec:fse}.) We will then perform a conventional analysis of
the $I$-$V$ exponent to look for evidence of $z\simeq5.9$ 
(Sec.~\ref{sec:convapp}.)
\subsubsection{Finite size effects}
\label{sec:fse}
Repaci {\it et al.}\cite{lobb96} have been suggested that their 
YBCO mono-layer data that has been shown here to 
scale so well with $z=5.9$ can be explained in a ``conventional''
manner by finite size effects, following similar suggestions regarding 
low current ohmic behavior of 
others.\cite{simkin97,vanderzant90,kim93} We 
investigate that possibility in this subsection, making
an explicit comparison between the dynamic scaling analysis 
and the conventional approach. As the reader will see, we 
find no evidence of finite size effects in any of the data that 
we examine.

This subsection is organized as follows. After a discussion 
of the principle  differences between the two scenarios, 
we look for evidence of finite size effects in the $I$-$V$
data from the YBCO mono-layer data\cite{lobb96} first using 
the conventional picture and then in the dynamic scaling 
analysis.

\centerline{\it Theory}

The primary difference between the finite size explanation 
and the dynamic scaling explanation of that data 
is the following: the 
former ascribes finite-size induced, low current ohmic behavior to $I$-$V$ 
isotherms with temperatures $T<T_{KTB}$; in the dynamic 
scaling explanation the low current ohmic behavior is associated 
with $T>T_{KTB}$ isotherms. The issue of course is the placement 
of the transition temperature. Because the scaling analysis 
indicates that $\alpha(T_{KTB})\simeq6.8$ and the conventional 
analysis assumes $\alpha(T_{KTB})=3$, $T_{KTB}$ will be 
at a lower temperature in the scaling analysis scenario.

We first examine the nature of the ohmic to non-ohmic crossover 
in the finite-size-effect scenario. Recall that with 
finite size effects, there is always a density of free 
vortices. [See Eq.~(\ref{density}).] This means that at 
small currents where one is probing large length scales, 
the free vortices will dominate the resistance and the 
$I$-$V$ characteristics will be ohmic. As one probes 
smaller length scales by increasing the current  ($r_c\ll L_{fs}$), 
current-induced vortex unbinding will start to dominate 
the resistance and the $I$-$V$ curves will become non-ohmic.\cite{blatter94} 
Because either the system size or the 2D penetration depth 
can be on the order of 1-10 microns, one would expect 
finite size effects to have an influence. 

We will now use the conventional picture to make the above 
discussion more quantitative for $T<T_{KTB}$. We will show
that the $I$-$V$ curves are ohmic when $r_c\gg L_{fs}$ and
non-ohmic when $r_c\ll L_{fs}$. With finite size effects,
the energy of a vortex pair is generalized from 
Eq.~(\ref{pairenergy}) to a potential that is logarithmic for
$R\ll L_{fs}$ and approaching a constant for $R\gg L_{fs}$.\cite{pearl64} 
To approximate this behavior, we use 
\begin{equation}
\label{pairenergyFS}
E(R)=q^2[ \ln (R/\xi_0)-\ln (R/L_{fs}+1)]-JR/\xi_0 +2E_c
\end{equation}
where $q^2=\pi n^{2D}_s\hbar^2/2m$ and $J=\pi\hbar I\xi_0  d/e A$ 
is a current with dimensions of energy. It is the second 
term on the right hand side of the equation that causes the 
zero-current ($J=0$) pair energy to approach a constant as 
$1/R$ for $R>L_{fs}$. With this term, E(R) no longer peaks 
at $r_c$. Rather, it peaks at
\begin{equation}
\label{rcFS}
r_c^{fs}=-\frac{L_{fs}}{2}+\frac{\sqrt{L_{fs}^2/4+L_{fs}\xi_0 q^2/J}}{2}.
\end{equation}
There are two limits to this equation. The large current limit 
$J\gg q^2\xi_0/L_{fs}$ can be rewritten as $r_c \ll L_{fs}$ while 
the small current limit can be expressed, $r_c \gg L_{fs}$. In 
the large current limit where one is probing small lengths, 
$r_c^{fs}$ approaches $r_c$ while at small currents 
($J\ll q^2\xi_0/L_{fs}$), $r_c^{fs}=\sqrt{L_{fs}\xi_0 q^2/J}.$ 
Remember that it is the value of $E(R)$ at $r_c^{fs}$ (i.e., 
the height of the barrier) that determines the density of free 
vortices. In one limit ($r_c \ll L_{fs}$), 
\begin{equation}
\label{rclargeJ}
E(r_c^{fs})=q^2[\ln(q^2/J)-1].
\end{equation}
On the other hand,
\begin{equation}
\label{rcsmallJ}
E(r_c^{fs})=q^2\ln(L_{fs}/\xi_0) 
\end{equation}
for $r_c\gg L_{fs}$. So, for $T<T_{KTB}$, the $I$-$V$'s are 
ohmic at small currents since $E(r_c^{fs})$ does not depend on 
current. At large currents, $E(r_c^{fs})$ depends upon the current 
in the traditional way, and one finds the usual $I$-$V$ 
relationship: $V\propto I^\alpha$. The ohmic to non-ohmic 
crossover occurs when 
$I\sim4k_BT_{KTB} W e/\pi\hbar L_fs$ (i.e., $r_c\sim L_{fs}$.)

The $T>T_{KTB}$ effect of finite size on the transport behavior 
is a little more subtle than the low temperature effect because
another competing length scale is involved. Even in the 
absence of finite size effects for this temperature range, the $I$-$V$ 
curves cross over from thermally dominated ohmic behavior at small 
$I$ to current-induced non-ohmic behavior at large $I$. The 
current at which this crossover occurs 
depends upon the size of the largest vortex pairs, 
$\xi_+(T)$,\cite{simkin97} a quantity 
which is strongly temperature dependent. With finite size effects, since all 
currents that probe lengths greater than $L_{fs}$ are ohmic,
$L_{fs}$ competes with $\xi_+(T)$ yielding the following conditions: 
when $r_c\gg \min[L_{fs},\xi_+]$ there is ohmic behavior 
and non-ohmic when $r_c\ll \min[L_{fs},\xi_+]$. An important term
in our discussion is ``premature-ohmic" behavior, which denotes ohmic 
behavior at currents for which there would not be ohmic behavior in the
absence of finite size effects. It occurs when $L_{fs}<\xi_+(T)$ 
(and when the temperature is sufficiently close to $T_{KTB}$.)

It is the temperature dependence of these conditions (as well as the magnitude
of the crossover current) that mark
the signature of finite size effects for $T>T_{KTB}$. 
For $T\gtrsim T_{KTB}$, $\xi_+(T)$ is large and exceeds $L_{fs}$
so that the value of the current at which 
the isotherms cross from one behavior to another will depend upon 
$L_{fs}$ and not the size of the largest pairs. In this case, the
crossover current will depend only weakly on temperature. 
As $\xi_+(T)$ becomes smaller than $L_{fs}$ at temperatures 
further above the transition, it is $\xi_+(T)$ that sets sets the current
scale for the ohmic to non-ohmic crossover. So the crossover current
becomes strongly temperature dependent.

\centerline{\it Conventional check of finite size effects}

As mentioned above possible candidate for observing finite size 
effects is the mono-layer YBCO data of Ref.~\onlinecite{lobb96} 
because the $\lambda_{2D}\simeq 40 \mu$m. To 
determine the value of the current at which ``premature-ohmic" 
behavior should occur for this data, we use $r_c=L_{fs}=40\mu$m 
to solve for the current. Using $W=200\mu$m, we find 
$I_{crossover}\simeq 2.4 \mu$A, 
a value much smaller than the observed crossover current: 
$\sim 100 \mu$A for $T\simeq28$K. This would seem to indicate 
that $\lambda_{2D}$ is not responsible for the ohmic to non-ohmic 
crossover in these materials. A renormalization group study\cite{pierson99}
confirms that this remains the case after renormalization effects 
of $L_{fs}$ are accounted for. This is because the condition for 
observing the finite size effects is $r_c>L_{fs}$. Under renormalization,
$L_{fs}$ does shrink quickly but $r_c$ shrinks even more
quickly.\cite{pierson99}

The length scale that corresponds to the approximate crossover 
current for the Repaci {\it et al.}\cite{lobb96} data for 
temperatures around 28K is $1\mu$m. Because neither a 
field-induced vortex length or a pinning (disorder) length could 
correspond to this value, it seems unlikely that the behavior observed 
in Ref.~\onlinecite{lobb96} is due to finite size effects.

Not only is the magnitude of the crossover current inconsistent with finite
size effect but so also is the temperature dependence of this quantity. 
If it were $L_{fs}$ and not $\xi_+$ which determined the ohmic to 
non-ohmic crossover, the data would not collapse so well in the dynamic 
analysis since the temperature dependence of these two quantities 
are so different.

Another signature for finite size effects in the conventional 
picture is that the resistance will have an arrhenious temperature 
dependence: $R(T)=A_N \exp [B/K_BT]$ where $B$ is related to 
the energy of a free vortex and $A_N$ is a constant. [See 
Eq.~(\ref{density}).] We have fit this formula to the data, as 
shown in Fig.~\ref{resfit} (dotted line), but do not get 
a satisfactory result. 

Finally, a finite size analysis of Repaci {\it et al.}\cite{lobb96} 
data would indicate a $T_{KTB}$ in excess of 30K following the 
subsequent work of Herbert {\it et al.}\cite{herbert98} This is 
contrary to the mutual 
inductance data of Gasparov {\it et al.}\cite{gasparov98} on 
YBCO mono-layer films which probes the temperature at which the 
largest pairs unbind. (Remember, the size of the largest pairs 
is expected to decrease as $\xi_+(T)$ above $T_{KTB}$.) 
They find that vortex pairs of size $\sim0.018\mu$m 
unbind at a temperature of roughly 47.0K, vortex pairs of 
size $\sim0.78\mu$m unbind at a temperature of 
roughly 27.8K, and vortex pairs of size 
$\sim1.58\mu$m unbind at a temperature of roughly 25.5K. 
This trend is consistent with the value of $T_{KTB}$ that we
find ($17.6$K) for the YBCO mono-layer data of Ref.~\onlinecite{lobb96} 
using a dynamic scaling analysis. 

\centerline{\it Dynamic scaling check of finite size effects}

The signature of finite size effects in the dynamic scaling 
depends upon the temperature. For $T<T_{KTB}$, the scaled 
data should peal off the scaling curve to go ohmic (slope 1 
in Figs.~\ref{1scsc}-\ref{sfsc}) for currents less than 
$4k_BT_{KTB} W e/\pi\hbar L_{fs}$. For temperatures above 
the transition temperature, one would observe finite size 
effects only if $L_{fs}$ were shorter than the size of the 
largest vortex pairs.  And in that case, the scaled data 
would break from the scaling curve to become 
``prematurely-ohmic" at $I=4k_BT_{KTB} W e/\pi\hbar L_{fs}$. 
This is not observed in any of the scaled data in this paper. 
Such behavior was observed however for the BSCCO data of 
Ref.~\onlinecite{ammirata98}. (See Fig.~2 there.) In that case 
however, it seems more likely that this behavior is not due to 
finite size effects but to voltage sensitivity. The crossover to 
the premature-ohmic behavior occurs more rapidly than one would 
expect for finite size effects and also occurs at roughly the same 
voltage, which is near the voltage sensitivity limit. 

\centerline{\it Finite size effect discussion}

As mentioned above, there is not a ``true'' thermodynamic phase 
transition in superconductors because of the finite penetration 
depth. One could then ask why there is a critical isotherm 
at all. For example, in the Repaci {\it et al.}\cite{lobb96} data, 
the 17K isotherm is straight over nearly 9 decades of voltage. 
The answer to this is that the correlation length at finite current 
does not become infinite (and is not longer than $L_{fs}$) even though 
it is very close to $T_c$. This is 
apparent from Eq.~(\ref{xiI}) where it is seen that the correlation 
length decreases as the reciprocal of current. It is only when 
$\xi(I,T)\simeq L_{fs}$ that one would begin to see deviations in the 
critical isotherm.

To summarize this section, we find no evidence of finite size 
effects in any of the data that we examine. In principle however, 
finite size effects are inherent to superconductors and will manifest
themselves if the probing current is small enough. It is our 
opinion that none of the data sets that we examined went to
currents small enough to detect finite size effects. 

\subsubsection{Conventional Approach}
\label{sec:convapp}

\begin{figure}
\centerline{
\epsfig{file=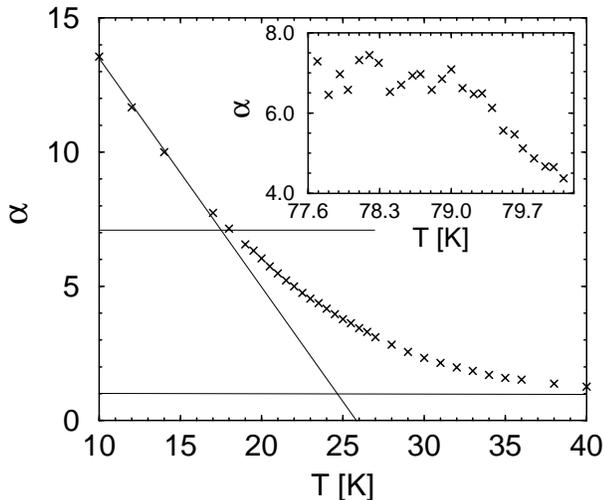,angle=-90,width=3.4in}}
\caption{$\alpha(T)$ data at a fixed current ($I=0.7$mA) for an
YBCO mono-layer taken from the $d[\log V]/d[\log I]$ data in Figure 3 of
Ref.~\protect\onlinecite{lobb96}. Also plotted are a linear fit to the data
less than the transition temperature and the lines $\alpha(T)=7.1$ and
$\alpha(T)=1.0$ used to determine the parameters $T_{c0}$, $z$, and
$\epsilon_c$. [INSET: $\alpha(T)$ data at a fixed current ($I\sim4.0$mA) for
the BSCCO film of Ref.~\protect\onlinecite{ammirata98}.]
}
\label{alphaT}
\end{figure}

We now examine the Repaci {\it et al.} data with a conventional
approach. In Figure \ref{alphaT}, we plot the $I$-$V$ exponent 
($V\propto I^{\alpha(T)}$) at a fixed current ($I=0.7$mA) as 
determined from the $d[\log V]/d[\log I]$ data in Figure 3 
of Ref.~\onlinecite{lobb96}.
(The error bars were determined from that figure and from fits to the 
$I$-$V$ curves, and are only shown for the near-linear region. 
The $\alpha(T<17$K) data also came from power-law fits to the 
digitized $I$-$V$ data from Fig.~2 of that reference.) As one 
can see, there are no features at $\alpha\sim3$ that would 
suggest a phase transition, as originally pointed out by those authors. 
An interpretation of this data is difficult. A possible feature 
is a crossover from near-linear behavior of $\alpha(T)$ to nonlinear 
behavior near the value of $T_{KTB}$ (17.6K), obtained from the 
dynamic scaling behavior. Further, the value of $\alpha$ observed 
at this temperature produces an estimate for $z$ ($\simeq 7$) that 
is similar to that obtained from the scaling procedure. One can see 
that the $T>T_{KTB}$ behavior is concave up, contrary to the analytical
work of Ref.~\onlinecite{goldman83}, but more consistent with the 
simulational work of Refs.~\onlinecite{holmlund96}.

The values of $T_{KTB}$ and $T_{c0}$ determined from the standard analysis 
of $\alpha(T)$ with arbitrary $z$ are less definitive 
because one is not assuming a value of $\alpha(T_{KTB})$. To be consistent 
with the value of $T_{KTB}$ determined from the scaling analysis, we 
have chosen $T_{KTB}=17.6$K. The subsequent value of $T_{c0}$ determined from
a linear fit to the $\alpha(T<17.6$K) is $T_{c0}=24.8$K. (See 
Figure \ref{alphaT}.) $z$ was found to be $6.1\pm0.2$, consistent
with the dynamic scaling value. The renormalized dielectric constant also
has a reasonable value: $\epsilon_c=1.59$.

One possible explanation for the absence of a clear jump in the 
exponent $\alpha (T)$ in Figure \ref{alphaT} is the relatively 
short length scale it represents.  In order to obtain well-defined
and stable $\alpha (T)$ values (at the same reference current for all 
of the isotherms), it was necessary to choose a relatively large value 
for the reference current.  This current value corresponds to the  
length scale $1400\AA$ ($\ll L_{fs}$), which is distant from the 
desired $I\rightarrow 0$ limit. 

\begin{figure}
\centerline{
\epsfig{file=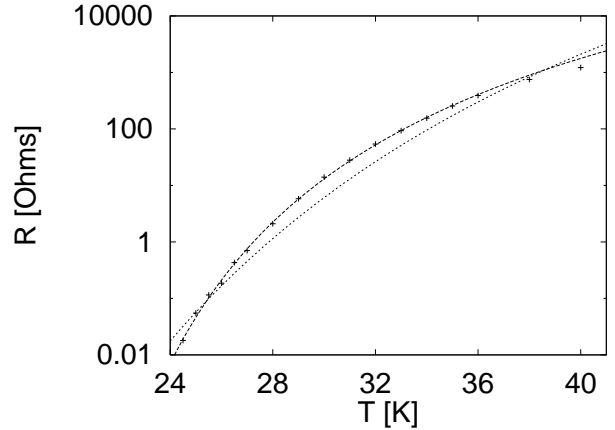,width=3.4in}}
\caption{The resistance data of Ref.~\protect\onlinecite{lobb96} and fits to 
Eq.~(\protect\ref{RT}) (dashed line) and Eq.~(\protect\ref{density}) (dotted line).
The latter equation, based on assumptions of finite size effects, does not
adequately describe the data.}
\label{resfit}
\end{figure}

The final step in the conventional approach is to examine $R(T)$. 
Eq.~(\ref{RT}) was used in place of Eq.~(\ref{RTmin}) to
fit the data since many of the isotherms are obtained in the
regime $T>T_{c0}=24.8$K, where Eq.~(\ref{RTmin}) is not valid. 
The dashed line in Figure \ref{resfit} is the fit using the parameters 
that optimized the scaling for Fig.~\ref{1scsc}b: 
$T_{KTB}=17.6$~K, $b=7.79$, and $z=5.9\pm0.3$.  Clearly, 
the data is more consistent with KTB critical behavior 
than finite size effects discussed  in the previous section. 

We repeat this analysis for the BSCCO sample ($1000 \AA$ thick) of 
Ref.~\onlinecite{ammirata98}. The $\alpha(T)$ determined for a constant
current is shown in the inset of Fig.~\ref{alphaT}. As one can see, this 
data is more noisy and covers a much smaller temperature range than
that in Fig.~\ref{alphaT}, thereby precluding a complete 
conventional analysis. So, while the value of $\alpha$ at $T_{KTB}=78.87$K 
is consistent with that determined from the dynamic scaling analysis
($6.6$) and $\alpha$ seems to change behavior at that 
temperature, one could not claim these observations as evidence 
for $\alpha(T_{KTB})=6.6$. However, an 
important observation can be made by comparing this inset to 
the inset of Fig.~1 of Ref.~\onlinecite{ammirata98}, which shows 
$\alpha(T)$ determined from the same data, but for a constant voltage.
As one can see, $\alpha(T)$ decreases much more rapidly for
a constant voltage than for a constant current, reinforcing our 
claim in Sec.~\ref{sec:alpha3to1} that a rapid decrease in $\alpha(T)$ 
could be an artifact of using the constant voltage $\alpha(T)$ data.

\subsection{Theoretical Considerations}
\label{sec:Theory}

The primary degrees of freedom associated with the
KTB phase transition are vortices.  The dynamic behavior
should therefore be dissipative.  Specifically, it has been argued\cite{ffh}
that superconducting dynamics in zero field may be purely relaxational 
(model A\cite{hohenberg}) for any dimension,
with a diffusive exponent, $z\leq 2$.  This interpretation is consistent
with the conventional treatment of KTB dynamics.
However, the present scaling analysis of $I$-$V$ and 
$\dot Q$-$\Delta T$ data from SC's, JJA's and SF's indicates that
$z\simeq 5.6$, a result consistent with sub-diffusive dynamics. 
Here we mention some
possible explanations for this large value of the dynamic exponent.

Pinning is known to play a crucial role in the large
values of $z$ ($\gg 2$) observed in
vortex glass phenomena in experiments\cite{koch89,olsson91,moloni} and
Monte Carlo simulations\cite{lidmar,leestroud,jensen} in high temperature
superconductors.  However, we do not believe that pinning can 
explain the surprising values of $z$ obtained in
the present 2D analysis, at zero field.  The reason is that our result, 
$z \simeq 5.6$ is obtained from very distinct systems:  
superconductors, JJA's, and superfluids.  For superfluids, 
in particular, pinning effects should be
negligible.  A pinning explanation therefore appears inconsistent with 
the universal nature of $z$.

Collective excitations, such as vortex density waves,\cite{hebboul99} may 
mediate the observed dynamic behavior.  This is more likely to be true 
if the vortices cannot exit the sample easily, because of surface barriers.
We can determine the dynamical critical exponent for this behavior as follows.
Based upon the Coulomb gas analogy, the vortex plasma frequency is given by
$\omega \propto \sqrt{n}$, where $n\sim \xi^{-d}$ is the vortex 
density.  Using $\tau\propto\xi^z$, we find $z=1.5$ in three
dimensions and $z=1.0$ in two dimensions.  Vortex plasmons therefore cannot 
explain the large values of $z$.

Another possibility is that the suppositions leading to Eq.~(\ref{vD}) are
incorrect. 
To arrive at that equation, only two forces are included, a viscous force and
the Lorentz force. Perhaps, with the inclusion of other forces (e.g. surface
barrier forces,\cite{fuchs98}) an explanation for $z=5.6$ can be found. 

We believe the most likely explanation for large values of $z$
lies in correlated vortex motion, described as 
``partner transfer,''\cite{bormann} or ``collaborative
dissociation".\cite{ammirata98} These suggest mechanisms whereby bound 
vortex pairs do not simply dissociate into free
vortices.  Instead, the process is mediated by neighboring vortex pairs, 
in terms of consecutive
recombination--dissociation events.  Further work is required to 
confirm this model.

\section{Summary}
\label{sec:summary}

As stated in the introductory paragraph, the dynamic scaling approach 
presented here is different than most previous studies of dynamics in
2D SC's, JJA's, and SF's in that this approach allows one to vary 
$z$ to optimize the analysis. In the 
``conventional'' approach, the value of $z$ is implicitly taken to 
be two. By using the dynamic scaling analysis and allowing the value of $z$
to vary, we have presented evidence which suggests non-diffusive behavior.

Via a dynamic scaling analysis of transport data from SC's, JJA's, and 
SF's, we find $z\simeq 5.6\pm0.3$, contrary to the value assumed but not 
tested in previous reports. This analysis seems convincing in that 
the collapse is excellent in many data sets and the value of $z$ is 
robust from system to system and material to material. 

The results of the dynamic scaling analysis also go against the many
studies consistent with $z=2$. 
We have included in this work a discussion of those ``conventional'' 
approaches to studying the dynamics of the KTB transition. Like the dynamic
scaling 
analysis, we find that this approach also has its drawbacks. Perhaps 
the most important is that the experiments do not yet have the 
sensitivity to actually observe the predicted jump in the $I$-$V$'s 
in the $I\rightarrow0$ limit which we estimated to be at $10^{-11}$A 
for a particular sample.\cite{lobb96} Another drawback is that these 
approaches do not vary $z$ to optimize the fits. Furthermore, the most 
common method of ``verifying'' KTB behavior [which is to obtain 
$T_{KTB}$ from the condition $\alpha(T_{KTB})=3$ and then to do a three 
parameter fit to Eq.~(\ref{RTmin}),] does 
not pose constraints tight enough to prove $z=2$.  Our analysis of 
the evidence for $z=2$ raises the following questions:
\begin{itemize}
\item Why has no scaling of zero-field $I$-$V$ data with $z=2$ been realized;
\item Why do critical isotherms have a much larger value of $\alpha$ 
than the value consistent with $z=2$;
\item If finite size effects are present, why does the ohmic to non-ohmic
crossover
not coincide with $r_c=L_{fs}$.
\end{itemize} 

We also compared directly the conventional approach and the
dynamic scaling approach for data from a particular sample: 
the YBCO mono-layer data of Ref.~\onlinecite{lobb96}. In that 
reference, those authors found that the conventional approach
can not explain their data. In Section \ref{sec:fse}, 
we found that an incorporation of the finite size effects into
the conventional approach is also not consistent with their data. Further
in Section \ref{sec:convapp}, we saw that a conventional analysis of the
$I$-$V$ exponent was consistent with $z\sim5.6$.
A dynamic scaling analysis of their data however resulted in a 
beautiful collapse, as shown in Fig.~\ref{1scsc}b.

The primary purpose of this paper is to convey that the question 
of the value of $z$ in these systems is still an open one, despite 
the conventional wisdom that $z=2$. We believe 
that more study is needed. In particular, more data on all systems, 
especially JJA and SF is needed and over wider temperatures and 
current regions (or $\dot Q$ regions for SF's). The impressive data 
of Repaci {\it et al.}\cite{lobb96} sets a good standard. Not only should
dynamic 
scaling analysis be tried on this data but so too should comprehensive 
``conventional" studies like those of Fiory, Hebard and 
Glaberson.\cite{fiory83} (By ``comprehensive'', we mean going beyond
just the usual $\alpha(T_{KTB})=3$ and $R(T)$ measurements.) 
Of special importance would 
be a measurement of  $q^2$ using static kinetic inductance data 
(in the appropriate frequency range) and the dynamic $I$-$V$ 
exponent $\alpha$. If the conventional theories are valid after 
generalization to a general $z$, then $\alpha-1=[z/2]q^2$. (Fiory 
{\it et al.} did do such a measurement but found agreement
only over $30$mK.) Allowing the value of $z$ in the 
conventional theory would also be a useful exercise.

The authors gratefully acknowledge conversations with S.~M.~Girvin,
L.~I.~Glazman, 
B.~I.~Halperin, A.~Hebard, J.~M.~Kosterlitz, C.~J.~Lobb, P.~Minnhagen,
P.~Muzikar, and S.~Teitel. 
This work was supported by the Midwest Superconductivity Consortium through 
D.O.E.~Contract No. DE-FG02-90ER45427 and by NSF Grant No.~DMR 95-01272. 
Acknowlegement (SWP) is made to the donors of The Petroleum Research Fund, 
administered by the ACS, for support of this research.

\end{document}